\documentstyle[10pt,epsf,epsfig,hangcaption,xspace,amssymb,amsfonts,amsmath,amsthm,cite,
dp_delphititle,lineno]{dp_delphi}
\makeindex
\def\DpPaperGroup{EP}
\def\DpPaperRef{2002-083}
\def\DpDate{16 October 2002}
\def\DpAuthors{DELPHI Collaboration}
\def\DpTitle{{
Search for supersymmetric particles in light gravitino scenarios and
sleptons NLSP}}
\def\DpSubmit{(Accepted by Eur. Phys. J. C)}
\def\DpComment{ }
\def\DpEMail{}

\newcommand{\Zqq}{\mbox{$e^+e^-~\rightarrow\,q\overline{q}$}}
\newcommand{\Chipm}{\tilde{\chi}^{\pm}}

\newcommand{\Chip}{\tilde{\chi}^{+}}
\newcommand{\Chim}{\tilde{\chi}^{-}}
\newcommand{\Chiz}{\tilde{\chi}^{0}}
\newcommand{\sGra}{\tilde{G}}
\newcommand{\sLep}{\tilde{l}}

\newcommand{\sLepR}{{$\tilde{l}_R$}}

\newcommand{\sTau}{{$\tilde{\tau}$}}

\newcommand{\stau} {$\tilde{\tau}$}
\newcommand{\staone}{$\tilde{\tau}_1$}
\newcommand{\stuno} {$\tilde{\tau}_1$}
\newcommand{\staur} {$\tilde{\tau}_R$}
\newcommand{\staul} {$\tilde{\tau}_L$}
\newcommand{\selr} {$\tilde{e}_R$}
\newcommand{\smur} {$\tilde{\mu}_R$}
\newcommand{\slep} {$\tilde{l}$}
\newcommand{\slepr} {$\tilde{l}_R$}
\newcommand{\nuno} {$\tilde{\chi}^0_1$}

\newcommand{\lmean} {$\hat{L}$}
\newcommand{\grav} {$\tilde{G}$}
\newcommand{\mgrav} {$m_{\tilde{G}}$}
\newcommand{\mstau}{$m_{\tilde{\tau}}$}

\newcommand{\ra} {\rightarrow}
\newcommand{\eeto} {\mbox{$ {\mathrm e}^+ {\mathrm e}^-\! \ra\ $}}
\newcommand{\ee} {\mbox{$ {\mathrm e}^+ {\mathrm e}^-$}}
\newcommand{\qqbar} {$q\bar{q}$}

\newcommand{\GeV} {~\mbox{${\mathrm{GeV}}$}}
\newcommand{\eeggg} {e^+e^-\rightarrow \gamma\gamma(\gamma)}
\newcommand{\eeSg} {e^+e^-\rightarrow \phi \gamma}
\newcommand{\TeVcc} {~\mbox{${\mathrm{TeV}}/{\mathrm{c}}^2$}}
\newcommand{\GeVcc} {~\mbox{${\mathrm{GeV}}/{\mathrm{c}}^2$}}
\newcommand{\eVcc} {~\mbox{${\mathrm{eV}}/{\mathrm{c}}^2$}}
\newcommand{\MeVc} {~\mbox{$ {\mathrm{MeV}}/ {\mathrm{c}} $}}

\newcommand{\GeVc} {~\mbox{$ {\mathrm{GeV}}/{\mathrm{c}} $}}
\newcommand{\TeV} {~\mbox{$ {\mathrm{TeV}} $}}
\newcommand{\keVcc} {~\mbox{$ {\mathrm{keV}}/{\mathrm{c}}^2 $}}
\newcommand{\etal} {\mbox{\it et al.}}
\def\NPB#1#2#3{{\rm Nucl.~Phys.} {\bf{B#1}} (19#2) #3}
\def\PLB#1#2#3{{\rm Phys.~Lett.} {\bf{B#1}} (19#2) #3}
\def\PRD#1#2#3{{\rm Phys.~Rev.} {\bf{D#1}} (19#2) #3}
\def\PRL#1#2#3{{\rm Phys.~Rev.~Lett.} {\bf{#1}} (19#2) #3}
\def\ZPC#1#2#3{{\rm Z.~Phys.} {\bf C#1} (19#2) #3}

\def\PR#1#2#3{{\rm Phys.~Rep.} {\bf#1} (19#2) #3}

\def\NIMA#1#2#3{{\rm Nucl.~Instr.~and~Meth.} {\bf#1} (19#2) #3}
\def\CPC#1#2#3{{\rm Comp.~Phys.~Comm.} {\bf#1} (19#2) #3}
\begin{document}
\makeatletter
\newcount\@tempcntc
\def\@citex[#1]#2{\if@filesw\immediate\write\@auxout{\string\citation{#2}}\fi
  \@tempcnta\z@\@tempcntb\m@ne\def\@citea{}\@cite{\@for\@citeb:=#2\do
    {\@ifundefined
       {b@\@citeb}{\@citeo\@tempcntb\m@ne\@citea\def\@citea{,}{\bf ?}\@warning
       {Citation `\@citeb' on page \thepage \space undefined}}%
    {\setbox\z@\hbox{\global\@tempcntc0\csname b@\@citeb\endcsname\relax}%
     \ifnum\@tempcntc=\z@ \@citeo\@tempcntb\m@ne
       \@citea\def\@citea{,}\hbox{\csname b@\@citeb\endcsname}%
     \else
      \advance\@tempcntb\@ne
      \ifnum\@tempcntb=\@tempcntc
      \else\advance\@tempcntb\m@ne\@citeo
      \@tempcnta\@tempcntc\@tempcntb\@tempcntc\fi\fi}}\@citeo}{#1}}
\def\@citeo{\ifnum\@tempcnta>\@tempcntb\else\@citea\def\@citea{,}%
  \ifnum\@tempcnta=\@tempcntb\the\@tempcnta\else
   {\advance\@tempcnta\@ne\ifnum\@tempcnta=\@tempcntb \else \def\@citea{--}\fi
    \advance\@tempcnta\m@ne\the\@tempcnta\@citea\the\@tempcntb}\fi\fi}
 
\makeatother
\begin{titlepage}
\pagenumbering{roman}
\CERNpreprint{\DpPaperGroup}{\DpPaperRef} 
\date{{\small\DpDate}} 
\title{\DpTitle} 
\address{\DpAuthors} 
\begin{shortabs} 
\noindent
A search for sleptons, neutralinos, charginos,
sgoldstinos and heavy stable charged sleptons in the context of scenarios
where the lightest su\-per\-sym\-me\-tric particle is the gravitino, is presented.  
Data collected during 2000 with the DELPHI detector at centre-of-mass
energies from 204 to 208 GeV were analysed and combined with all the
data collected from 1995 to 1999 at lower energies.  No evidence 
for the production of sleptons, neutralinos and charginos has been found, 
therefore
new limits on the mass of these supersymmetric particles and on the model parameter
space are set. The search for
heavy stable charged sleptons also updates the 
stable sleptons mass limit.
The absence of evidence for sgoldstino production allows limits to be
set on its mass and on the scale of supersymmetry breaking.
\end{shortabs}
\vfill
\begin{center}
\DpSubmit \ \\ 
\DpComment \ \\
\DpEMail \ \\
\end{center}
\vfill
\clearpage
\headsep 10.0pt
\addtolength{\textheight}{10mm}
\addtolength{\footskip}{-5mm}
\begingroup
%
\newcommand{\DpName}[2]{\hbox{#1$^{\ref{#2}}$},\hfill}
\newcommand{\DpNameTwo}[3]{\hbox{#1$^{\ref{#2},\ref{#3}}$},\hfill}
\newcommand{\DpNameThree}[4]{\hbox{#1$^{\ref{#2},\ref{#3},\ref{#4}}$},\hfill}
\newskip\Bigfill \Bigfill = 0pt plus 1000fill
\newcommand{\DpNameLast}[2]{\hbox{#1$^{\ref{#2}}$}\hspace{\Bigfill}}
%
\footnotesize
\noindent
\DpName{J.Abdallah}{LPNHE}
\DpName{P.Abreu}{LIP}
\DpName{W.Adam}{VIENNA}
\DpName{P.Adzic}{DEMOKRITOS}
\DpName{T.Albrecht}{KARLSRUHE}
\DpName{T.Alderweireld}{AIM}
\DpName{R.Alemany-Fernandez}{CERN}
\DpName{T.Allmendinger}{KARLSRUHE}
\DpName{P.P.Allport}{LIVERPOOL}
\DpName{U.Amaldi}{MILANO2}
% ...
\normalsize
\endgroup
\titlefoot{Department of Physics and Astronomy, Iowa State
     University, Ames IA 50011-3160, USA
    \label{AMES}}
\titlefoot{Physics Department, Universiteit Antwerpen,
     Universiteitsplein 1, B-2610 Antwerpen, Belgium \\
     \indent~~and IIHE, ULB-VUB,
     Pleinlaan 2, B-1050 Brussels, Belgium \\
     \indent~~and Facult\'e des Sciences,
     Univ. de l'Etat Mons, Av. Maistriau 19, B-7000 Mons, Belgium
    \label{AIM}}
\titlefoot{Physics Laboratory, University of Athens, Solonos Str.
     104, GR-10680 Athens, Greece
    \label{ATHENS}}
\titlefoot{Department of Physics, University of Bergen,
     All\'egaten 55, NO-5007 Bergen, Norway
    \label{BERGEN}}
\titlefoot{Dipartimento di Fisica, Universit\`a di Bologna and INFN,
     Via Irnerio 46, IT-40126 Bologna, Italy
    \label{BOLOGNA}}
\titlefoot{Centro Brasileiro de Pesquisas F\'{\i}sicas, rua Xavier Sigaud 150,
     BR-22290 Rio de Janeiro, Brazil \\
     \indent~~and Depto. de F\'{\i}sica, Pont. Univ. Cat\'olica,
     C.P. 38071 BR-22453 Rio de Janeiro, Brazil \\
     \indent~~and Inst. de F\'{\i}sica, Univ. Estadual do Rio de Janeiro,
     rua S\~{a}o Francisco Xavier 524, Rio de Janeiro, Brazil
    \label{BRASIL}}
\titlefoot{Coll\`ege de France, Lab. de Physique Corpusculaire, IN2P3-CNRS,
     FR-75231 Paris Cedex 05, France
    \label{CDF}}
\titlefoot{CERN, CH-1211 Geneva 23, Switzerland
    \label{CERN}}
\titlefoot{Institut de Recherches Subatomiques, IN2P3 - CNRS/ULP - BP20,
     FR-67037 Strasbourg Cedex, France
    \label{CRN}}
\titlefoot{Now at DESY-Zeuthen, Platanenallee 6, D-15735 Zeuthen, Germany
    \label{DESY}}
\titlefoot{Institute of Nuclear Physics, N.C.S.R. Demokritos,
     P.O. Box 60228, GR-15310 Athens, Greece
    \label{DEMOKRITOS}}
\titlefoot{FZU, Inst. of Phys. of the C.A.S. High Energy Physics Division,
     Na Slovance 2, CZ-180 40, Praha 8, Czech Republic
    \label{FZU}}
\titlefoot{Dipartimento di Fisica, Universit\`a di Genova and INFN,
     Via Dodecaneso 33, IT-16146 Genova, Italy
    \label{GENOVA}}
\titlefoot{Institut des Sciences Nucl\'eaires, IN2P3-CNRS, Universit\'e
     de Grenoble 1, FR-38026 Grenoble Cedex, France
    \label{GRENOBLE}}
\titlefoot{Helsinki Institute of Physics, HIP,
     P.O. Box 9, FI-00014 Helsinki, Finland
    \label{HELSINKI}}
\titlefoot{Joint Institute for Nuclear Research, Dubna, Head Post
     Office, P.O. Box 79, RU-101 000 Moscow, Russian Federation
    \label{JINR}}
\titlefoot{Institut f\"ur Experimentelle Kernphysik,
     Universit\"at Karlsruhe, Postfach 6980, DE-76128 Karlsruhe,
     Germany
    \label{KARLSRUHE}}
\titlefoot{Institute of Nuclear Physics,Ul. Kawiory 26a,
     PL-30055 Krakow, Poland
    \label{KRAKOW1}}
\titlefoot{Faculty of Physics and Nuclear Techniques, University of Mining
     and Metallurgy, PL-30055 Krakow, Poland
    \label{KRAKOW2}}
\titlefoot{Universit\'e de Paris-Sud, Lab. de l'Acc\'el\'erateur
     Lin\'eaire, IN2P3-CNRS, B\^{a}t. 200, FR-91405 Orsay Cedex, France
    \label{LAL}}
\titlefoot{School of Physics and Chemistry, University of Lancaster,
     Lancaster LA1 4YB, UK
    \label{LANCASTER}}
\titlefoot{LIP, IST, FCUL - Av. Elias Garcia, 14-$1^{o}$,
     PT-1000 Lisboa Codex, Portugal
    \label{LIP}}
\titlefoot{Department of Physics, University of Liverpool, P.O.
     Box 147, Liverpool L69 3BX, UK
    \label{LIVERPOOL}}
\titlefoot{LPNHE, IN2P3-CNRS, Univ.~Paris VI et VII, Tour 33 (RdC),
     4 place Jussieu, FR-75252 Paris Cedex 05, France
    \label{LPNHE}}
\titlefoot{Department of Physics, University of Lund,
     S\"olvegatan 14, SE-223 63 Lund, Sweden
    \label{LUND}}
\titlefoot{Universit\'e Claude Bernard de Lyon, IPNL, IN2P3-CNRS,
     FR-69622 Villeurbanne Cedex, France
    \label{LYON}}
\titlefoot{Dipartimento di Fisica, Universit\`a di Milano and INFN-MILANO,
     Via Celoria 16, IT-20133 Milan, Italy
    \label{MILANO}}
\titlefoot{Dipartimento di Fisica, Univ. di Milano-Bicocca and
     INFN-MILANO, Piazza della Scienza 2, IT-20126 Milan, Italy
    \label{MILANO2}}
\titlefoot{IPNP of MFF, Charles Univ., Areal MFF,
     V Holesovickach 2, CZ-180 00, Praha 8, Czech Republic
    \label{NC}}
\titlefoot{NIKHEF, Postbus 41882, NL-1009 DB
     Amsterdam, The Netherlands
    \label{NIKHEF}}
\titlefoot{National Technical University, Physics Department,
     Zografou Campus, GR-15773 Athens, Greece
    \label{NTU-ATHENS}}
\titlefoot{Physics Department, University of Oslo, Blindern,
     NO-0316 Oslo, Norway
    \label{OSLO}}
\titlefoot{Dpto. Fisica, Univ. Oviedo, Avda. Calvo Sotelo
     s/n, ES-33007 Oviedo, Spain
    \label{OVIEDO}}
\titlefoot{Department of Physics, University of Oxford,
     Keble Road, Oxford OX1 3RH, UK
    \label{OXFORD}}
\titlefoot{Dipartimento di Fisica, Universit\`a di Padova and
     INFN, Via Marzolo 8, IT-35131 Padua, Italy
    \label{PADOVA}}
\titlefoot{Rutherford Appleton Laboratory, Chilton, Didcot
     OX11 OQX, UK
    \label{RAL}}
\titlefoot{Dipartimento di Fisica, Universit\`a di Roma II and
     INFN, Tor Vergata, IT-00173 Rome, Italy
    \label{ROMA2}}
\titlefoot{Dipartimento di Fisica, Universit\`a di Roma III and
     INFN, Via della Vasca Navale 84, IT-00146 Rome, Italy
    \label{ROMA3}}
\titlefoot{DAPNIA/Service de Physique des Particules,
     CEA-Saclay, FR-91191 Gif-sur-Yvette Cedex, France
    \label{SACLAY}}
\titlefoot{Instituto de Fisica de Cantabria (CSIC-UC), Avda.
     los Castros s/n, ES-39006 Santander, Spain
    \label{SANTANDER}}
\titlefoot{Inst. for High Energy Physics, Serpukov
     P.O. Box 35, Protvino, (Moscow Region), Russian Federation
    \label{SERPUKHOV}}
\titlefoot{J. Stefan Institute, Jamova 39, SI-1000 Ljubljana, Slovenia
     and Laboratory for Astroparticle Physics,\\
     \indent~~Nova Gorica Polytechnic, Kostanjeviska 16a, SI-5000 Nova Gorica, Slovenia, \\
     \indent~~and Department of Physics, University of Ljubljana,
     SI-1000 Ljubljana, Slovenia
    \label{SLOVENIJA}}
\titlefoot{Fysikum, Stockholm University,
     Box 6730, SE-113 85 Stockholm, Sweden
    \label{STOCKHOLM}}
\titlefoot{Dipartimento di Fisica Sperimentale, Universit\`a di
     Torino and INFN, Via P. Giuria 1, IT-10125 Turin, Italy
    \label{TORINO}}
\titlefoot{INFN,Sezione di Torino, and Dipartimento di Fisica Teorica,
     Universit\`a di Torino, Via P. Giuria 1,\\
     \indent~~IT-10125 Turin, Italy
    \label{TORINOTH}}
\titlefoot{Dipartimento di Fisica, Universit\`a di Trieste and
     INFN, Via A. Valerio 2, IT-34127 Trieste, Italy \\
     \indent~~and Istituto di Fisica, Universit\`a di Udine,
     IT-33100 Udine, Italy
    \label{TU}}
\titlefoot{Univ. Federal do Rio de Janeiro, C.P. 68528
     Cidade Univ., Ilha do Fund\~ao
     BR-21945-970 Rio de Janeiro, Brazil
    \label{UFRJ}}
\titlefoot{Department of Radiation Sciences, University of
     Uppsala, P.O. Box 535, SE-751 21 Uppsala, Sweden
    \label{UPPSALA}}
\titlefoot{IFIC, Valencia-CSIC, and D.F.A.M.N., U. de Valencia,
     Avda. Dr. Moliner 50, ES-46100 Burjassot (Valencia), Spain
    \label{VALENCIA}}
\titlefoot{Institut f\"ur Hochenergiephysik, \"Osterr. Akad.
     d. Wissensch., Nikolsdorfergasse 18, AT-1050 Vienna, Austria
    \label{VIENNA}}
\titlefoot{Inst. Nuclear Studies and University of Warsaw, Ul.
     Hoza 69, PL-00681 Warsaw, Poland
    \label{WARSZAWA}}
\titlefoot{Fachbereich Physik, University of Wuppertal, Postfach
     100 127, DE-42097 Wuppertal, Germany \\
\noindent
{$^\dagger$~deceased}
    \label{WUPPERTAL}}
\addtolength{\textheight}{-10mm}
\addtolength{\footskip}{5mm}
\clearpage
\headsep 30.0pt
\end{titlepage}
%
\pagenumbering{arabic} 
\setcounter{footnote}{0} %
\large
\section{Introduction}
\label{sec:intro}
In the year 2000 the Large Electron Positron Collider (LEP) at CERN finished
its ope\-ra\-tion achieving record energies of 204 to 208 GeV when the DELPHI
detector collected an integrated luminosity of \mbox{223.53 pb$^{-1}$}.
These data were analysed to update the searches for sleptons, neutralinos,
charginos, sgoldstinos  and heavy stable
charged particles~\cite{nuestro_papel_202,ref:delphisgold} 
in the context of gauge mediated
supersymmetry breaking (GMSB) models. 

Supersymmetry (SUSY) is usually assumed to be broken in a hidden sector of particles
and then communicated to the observable sector (where all the particles and
their superpartners lie) via gravitational interactions. 
An alternative possibility is that this mediation is performed
by Standard Model (SM) gauge interactions, leading to models of
gauge mediated supersymmetry breaking. In most current GMSB theoretical work
\cite{ref:dtw,ref:akm,ref:gr}, 
it is assumed that the hidden sector is coupled to a messenger
sector, which in turn couples to the visible sector through radiative
corrections with gauge-interaction strength. The primary motivation for GMSB
is that it naturally accommodates the experimentally observed absence of
flavour changing neutral currents due to the fact that gauge interactions are
flavour blind. In these models the scale of SUSY breaking ($\sqrt{F}$) can be
as low as about 10$^4$ or 10$^5$ GeV\footnote{In gravity mediated SUSY
  breaking models $\sqrt{F}~\sim$ 10$^{10}$ or 10$^{11}$ GeV.} in order to have
supersymmetric particle (sparticle) masses of the right order of magnitude
($\sim$ 100\GeVcc). 

The mass of the gravitino (\grav) is related to the scale of SUSY breaking
through the expression:
\begin{equation}
m_{\tilde{G}} \simeq 2.5 \times F / (100~{\rm TeV})^2~{\rm eV/c^2},
\end{equation}
therefore \mgrav~can be as low as few\eVcc. Consequently in these models
\grav~is the lightest supersymmetric particle (LSP) and all the other 
sparticles will decay into final states that include it.
Gravitino masses below 3$\cdot$10$^{-4}$\eVcc\ have been ruled out using
Tevatron 
data for the multijet final states~\cite{TeVlightGravLimit}. 
On the
other hand, hints from cosmology point in the direction of either a light
gravitino with mass below 1\keVcc\ or a heavy one with mass above
1\TeVcc~\cite{cosmology,wagner}. 

In GMSB models the entire minimal supersymmetric standard model (MSSM)
spectrum can be predicted in terms of the following parameters:
\begin{equation}
  F,~\Lambda, ~ M, ~ n, ~ \tan{\beta} ~ {\rm and} ~ {\rm sign}(\mu).
  \label{e:params}
\end{equation}
The most important parameter is $\Lambda$ (the effective SUSY breaking scale)
because it sets 
the overall mass scale of supersymmetric particles. $M$ is the messenger mass
scale. The number of messenger generations, $n$, is also very
important because it determines which sparticle is the next-to-lightest
supersymmetric particle (NLSP). For $n$ = 1 the NLSP is mainly the \nuno, 
and for $n\ge$ 2 it is 
one of the sleptons. The parameter tan$\beta$ is the ratio of the Higgs vacuum
expectation values, and sign($\mu$) is the sign of the Higgs sector mixing
parameter\footnote{The magnitude of $\mu$ is calculable from the other
  parameters in the model by imposing radiative electroweak symmetry
  breaking.}.

The coupling to the gravitino is very weak, therefore, all the superparticles 
other than the next-to-lightest supersymmetric particle undergo 
chain decay down to the NLSP which finally 
decays to the $\tilde{G}$. The mean decay length ($\hat{L}$) of the NLSP depends on
$m_{\tilde{G}}$~\cite{martin}. Namely, for $\tilde{l}\rightarrow l \tilde{G}$
decay:  
\begin{equation}
\hat{L} = 1.76 \times 10^{-3} \sqrt{\left (\frac{E_{\tilde{l}}}{m_{\tilde{l}}}
\right )^2-1}
\left ( \frac{m_{\tilde{l}}}{100 \, {\rm GeV/c}^2} \right )^{-5}
\left ( \frac{m_{\tilde{G}}}{1\, \rm eV/c^2}\right )^{2} \;\; {\mathrm cm}
\label{life}
\end{equation}
\noindent
where $m_{\tilde{l}}$ is the slepton mass and $E_{\tilde{l}}$  the slepton energy.
Therefore, the gravitino mass determines if the NLSP decays inside or outside
the detector, giving rise to very interesting topologies explored in this
paper.
For example, for $m_{\tilde{G}}~\lesssim~250$\eVcc ($\sqrt{F} \lesssim$
1000\TeV), the decay of a NLSP
with mass greater than for example 60\GeVcc~can take place
within the detector. This range of $\sqrt{F}$ is in fact consistent with
astrophysical and cosmological considerations \cite{Dinopoulos0,wagner}.
Figure \ref{fig.meandecay} shows the \stau~mean decay length 
as a function of the gravitino mass for different \stau~masses.

In this paper data were analysed within the two possible slepton NLSP
scenarios as discussed in the following. Depending on the magnitude of the mixing
in the third family between the left and right gauge eigenstates, \staur~and
\staul, there are two possible 
scenarios. If the mixing is large\footnote{In GMSB models large mixing 
occurs generally in regions of $\tan\beta\geq 1$0  or $|\mu|>$ 500\GeVcc.}, 
\stuno\ (the lighter mass eigenstate)
is the NLSP. However, if the mixing 
is negligible, \stuno\ is mainly right-handed~\cite{bartl}
and almost mass degenerate with the other sleptons. In this case, the
\selr\ and \smur\ three 
body decay (\slep $\rightarrow$ \stuno $\tau l$ with
\stuno $\rightarrow~\tau$ \grav), is very suppressed, and \selr~and
\smur~decay directly into $l$\grav. This scenario is called sleptons co-NLSP.

The signature for SUSY particle production within GMSB models
at LEP2 depends on the NLSP type and on its mean decay length, or 
equivalently, on the gravitino mass.
The NLSP could be pair produced directly, or other
sparticle production could lead to a cascade decay
into the NLSP.  The NLSP will decay into its non-SUSY partner and a \grav.
Taking into account all these factors, the following topologies can be 
expected: 
\begin{itemize}
\item For \mgrav~below a few eV/c$^2$, the NLSP decays
in the vicinity of its production point, before the tracking devices of the
detector, and different topologies can be expected. If the 
sleptons are pair produced the signature in the detector is the same as in
the search for sleptons in gravity-mediated supersymmetry breaking scenarios
(MSUGRA) with \nuno~LSP,  i.e. two acoplanar\footnote{Acoplanarity is defined as the complement of the
  angle between the projections of the two tracks onto the plane
  perpendicular to the beam.} leptons and missing energy~\cite{susana_paper}
since the slepton decays into a lepton and a neutralino that escapes
detection. This process is topologically equivalent to the pair production
of two sleptons and each of them decaying into a lepton and a gravitino (if
\mgrav~is below a few eV/c$^2$).
However, if neutralino pair production is 
kinematically allowed, the production cross-section can be larger than for
\slep~even if $m_{\tilde{\chi}^{0}_1}>m_{\tilde{l}}$ because of the $\beta^3$ suppression
factor of the scalar production cross-section. In this case the topology is
given by four leptons and missing energy since each $\Chiz_1$ decays into 
$\sLep l$, and the sleptons into $l\sGra$. 
The mass region that can be inspected using this search can be complemented
with the search  for lightest neutralino pair production when the neutralino
is the NLSP decaying into a photon and a gravitino~\cite{2gamma_189}.

\item For $m_{\sGra}$ between a few eV/c$^2$ and a few hundred eV/$c^2$ the NLSP has an
intermediate mean decay length and it would decay in flight in some part of the
detector volume. This  creates well defined secondary vertices or kinks when
the 
$\sLep$ is reconstructed by the tracking devices, or large impact parameter
tracks if it is not.  

\item For gravitino masses above few hundred  eV/c$^2$ the NLSP would be
sufficiently long-lived to decay outside the detector giving rise to
 heavy stable charged particle signatures.
\end{itemize}

In the GMSB parameter space where the
$\Chiz_1$ is the NLSP, the chargino is always much heavier than 100
GeV/c$^2$~\cite{D0} and, therefore, cannot be produced at LEP. 
On the contrary, in the parameter space where the
$\sLep$ is the NLSP there are regions where the $\Chipm_1$ is light enough
to be produced~\cite{Cheung}. In this case the topology for \mgrav~below a
few eV/c$^2$ is again two acoplanar leptons and 
missing energy since each $\Chipm_1$ decays into $\sLep \nu$ and each $\sLep$
into $l\sGra$. For $m_{\sGra}$ between a few eV/c$^2$ and a few hundred
eV/c$^2$ the topologies are  kinks or large impact parameter tracks
and, for gravitino masses above a few hundred  eV/c$^2$, heavy stable charged
particle signatures are expected.

In this paper, the update of the search for heavy stable charged
particles is also performed.  This kind of particle is predicted not only
in GMSB models but also in MSSM with a very
small amount of R-parity violation~\cite{Dreiner}, 
or with R-parity conservation 
if the mass difference between the LSP and the NLSP becomes very 
small (references in~\cite{degenerados}). 
The typical signature of these events is two 
massive particles traversing the detector which do not produce Cherenkov
radiation in DELPHI's Ring Imaging CHerenkov (RICH) detectors, but high
ionization losses in the 
Time Projection Chamber. Updated lower mass limits on
heavy stable charged particles, under the assumption that 
the LSP is a charged slepton, are presented. 

Recently it has been pointed out~\cite{ref:prz} that an appropriate 
theory must also contain the supersymmetric partner of the goldstino,
called the sgoldstino, which could be massive. 
In the minimal R-parity conserving model, as considered in  \cite{ref:prz},
the effective theory at the weak scale contains two neutral scalar states:
$S$, CP-even and $P$, CP-odd (from now on the two states will be labelled
 with
the generic symbol $\phi$ since the discussion applies to both of them). 
It must be pointed out that sgoldstinos have even R-parity,  
therefore  they are not necessarily produced in pairs and their decay chains 
do not necessarily contain the LSP.
The production of these supersymmetric particles may be relevant at LEP2
energies in light gravitino scenarios.
One of the most interesting production channels is
the process  $\eeSg$ which depends on the $\phi$ mass ($m_\phi$) and on
$\sqrt{F}$.
The most relevant $\phi$ decay modes  are $\phi\rightarrow \gamma \gamma$ and 
$\phi \rightarrow gg$. The corresponding branching ratios
depend on the gaugino masses $M_1,~M_2$ and $M_3$,
and the total width is  
$\Gamma\sim \Gamma(\phi\rightarrow \gamma \gamma)+\Gamma(\phi\rightarrow g g)$. 
In this paper two sets for these parameters are considered as suggested in \cite{ref:prz}; 
they are listed in Table \,\ref{tab:param}.
The total width for  a large interval of the parameter
space is  narrow (below a few GeV/c$^2$), except  for the region with small
$\sqrt{F}$ where the production cross-section is expected to be very
large. The two decay channels considered produce events with very different 
topologies. The channel $\phi\rightarrow \gamma \gamma$ gives 
events with three high energetic photons, one of which has monochromatic energy
$(E_{\gamma}=\frac{s-m^2_{\phi}}{2 \sqrt{s} })$ for a  
large fraction of the parameter space where  $\phi$ has a negligible
width. Despite  the lower $\phi$ decay branching ratio
(4 and 11$\%$ for the two sets of Table  \,\ref{tab:param}, respectively),
this final state is worth investigating because  the main background source is 
the QED process $\eeggg$, 
which is expected to be small if photons in the forward region are discarded. 
On the other hand, the channel $S\rightarrow  g g$ gives events with
one monochromatic photon (except for the region with small $\sqrt{F}$) and
two jets. An irreducible background from 
$e^+ e^- \rightarrow q \bar{q} \gamma$ events is associated to this topology
and therefore the signal must be searched for as an excess of events 
over the  background expectations for every mass hypothesis.
\begin{table}[h]
\begin{center}
\begin{tabular}{|c|c|c|c|c|c|}
\hline
 &$M_1$&$M_2$& $M_3$ &$BR(\phi\rightarrow \gamma \gamma)$&$BR(\phi\rightarrow g g)$   \\      
\hline
1)      & 200   & 300         & 400 & 4$\%$ &96$\%$ \\
\hline
2)      & 350   & 350         & 350 &11$\%$ &89$\%$ \\ 
\hline
\end{tabular}
\caption[]{Two choices for the gaugino 
mass parameters (in GeV/c$^2$) relevant for the sgoldstino production and
decay, and the corresponding branching ratios (BR) of the two  channels considered. The BR are almost independent of $m_\phi$ in the mass region below 200\GeVcc.}
\label{tab:param}
\end{center}
\vspace{-.5cm}
\end{table}

The list of GMSB signatures analysed in this paper is given in 
Table~\ref{t:gmsbtop}.

\begin{table}[h]
  \begin{center}
    \begin{tabular}{|l|l|l|l|}
      \hline
      Production & Decay mode & $\hat{L}$ & Expected topology  \\
      \hline
      \hline
      &  & $ << \ell_{detector}$ & Acoplanar leptons  \\
      ${\rm e^+e^-} \to \sLep \sLep$ & $\sLep \to l \sGra$   &  $\sim \ell_{detector}$ & Kinks and large impact parameters \\
      &  &  $ >> \ell_{detector}$ & Heavy stable charged particles   \\
      \hline
      ${\rm e^+e^-} \to \Chiz_1 \Chiz_1$ & $\Chiz_1 \to \sLep l \to l l \sGra$  &  $<< \ell_{detector}$ &  Four leptons \\
      \hline
      & &  $ << \ell_{detector}$ & Acoplanar leptons   \\
      ${\rm e^+e^-} \to \Chip_1 \Chim_1$ & $\Chip_1 \to \sLep^+
      \nu \to l^+ \sGra \nu$  &  $\sim \ell_{detector}$ & Kinks and large impact parameters \\
      &  & $ >> \ell_{detector}$ & Heavy stable charged particles  \\
      \hline
      ${\rm e^+e^-} \to \phi \gamma$ & $\phi \rightarrow \gamma\gamma$ & $ <<
      \ell_{detector}$ & 3 high energetic $\gamma$ \\
      & $\phi \rightarrow gg$ & $ << \ell_{detector}$ & 1 monochromatic
      $\gamma$ and 2 jets \\ \hline
    \end{tabular}
  \end{center}
    \caption[]{\label{t:gmsbtop}{Final state topologies studied in the different scenarios.\\}}
    \vspace{0.4cm}
\end{table}

The organization of the paper is as follows. 
A brief description of the DELPHI detector is presented in
section~\ref{sec.delphi}. The data samples are described in 
section~\ref{experimentalprocedure}. The different
selection criteria, the efficiencies and the number of events selected in
data and in the expected Standard Model background are reported in
section~\ref{dataselection}. Finally, the 
results are presented in section~\ref{sec:resultados} comprising
cross-section limits of the pair produced sparticles, lower mass limits and
limits on the GMSB model parameters.
%
%
\section{Detector description}
\label{sec.delphi}
DELPHI~was one of the four detectors operating at the LEP~collider from 1989
to 2000. 
It was designed as a general purpose detector for $e^+e^-$ physics
with special emphasis on precise
tracking and vertex determination and on powerful particle identification. 
A detailed
description of the DELPHI detector can be found in \cite{detector} and the
detector and trigger performance in \cite{performance,trigger}. 
Here only those components
relevant for the present analyses are discussed.

Charged particle tracks are reconstructed by a system of tracking chambers 
inside the 1.2~T solenoidal magnetic field: the Vertex Detector (VD), the Inner
Detector (ID), the Time Projection Chamber (TPC) and the Outer Detector
(OD) in the barrel region; two sets of planar drift chambers aligned
perpendicular to the beam axis (Forward Chambers A and B) measure tracks 
in the forward and backward directions.

For the data presented here, the VD consists of three cylindrical layers of 
silicon detectors, at radii
6.3~cm, 9.0~cm and 11.0~cm, and polar angle acceptance from 24$^\circ$ to 156$^\circ$. 
All three layers measure coordinates in the plane
transverse to the beam ($xy$), and at least two of the layers also measure
$z$ coordinates along the beam direction.  
The ID consists of a cylindrical drift chamber with inner radius 12~cm and
outer radius 22~cm, surrounded by 5 layers of straw tubes, having 
a polar acceptance between 15$^\circ$ and 165$^\circ$.
The TPC, the principal tracking device of DELPHI, consists of a 2.7~m long cylinder of
30~cm inner radius and 122~cm outer radius. Each
end-plate of the TPC is divided into 6 sectors with 192 sense wires
and 16 circular pad rows per sector. The wires help in charged particle
identification by measuring the specific energy loss
(dE/dx) and the pad rows are used for 3 dimensional
space-point reconstruction. 
The OD consists of 5 layers of drift cells at radii between 192~cm and
208~cm, covering polar angles between 43$^\circ$ and 137$^\circ$.

The electromagnetic calorimeters consist of a High Density Projection Chamber
(HPC) covering the polar angle region from 40$^\circ$ to 140$^\circ$ and, a
Forward ElectroMagnetic Calorimeter (FEMC) covering the polar angle regions from 
11$^\circ$ to 36$^\circ$ and from 144$^\circ$ to 169$^\circ$.  The Scintillator
TIle Calorimeter (STIC) extends the polar angle coverage down to 1.66$^\circ$
from the beam axis in both directions. 
The Hadron CALorimeter (HCAL) covers 98\% of the solid
angle. The muons which traverse the HCAL are recorded in a set of Muon Drift
Chambers placed in the barrel, forward and backward regions.

The Ring Imaging CHerenkov (RICH) detectors of DELPHI~provide charged particle 
identification in both the barrel (BRICH)
and forward (FRICH) regions. They contain two radiators of 
different refractive indices. The liquid radiator
is used for particle identification in the momentum range from 0.7 to 8
GeV/c. The gas radiator is used for particles with momentum range 
from 2.5 GeV/c to 25 GeV/c. 
  
%
%
\section{Data sample and event generators}
\label{experimentalprocedure}

The searches reported in this paper are based on data collected with the DELPHI detector
during 2000 at centre-of-mass energies from around 204 to 208 GeV.
The total integrated lu\-mi\-no\-si\-ty was 223.53~pb$^{-1}$.  
Table \ref{tab:lep2} summarises the energies
analysed and the integrated luminosities corresponding to each energy during 
the LEP2 period. 
\begin{table}[tbh]
\begin{center}
\begin{tabular}{|c|c|c|c|c|c|c|} \hline
Year & 1995 & 1996 & 1997 & 1998 & 1999 & 2000 \\ \hline
$\sqrt{s}$ (GeV)  & 130-136 & 161-172 & 183 & 189 & 192-202 & 204-208 \\ \hline
${\cal L}$ (pb$^{-1}$) & 11.9 & 19.6 & 54.0 & 158.0 & 228.2 & 223.5 \\ \hline
\end{tabular}
\end{center}
\caption{Centre-of-mass energies analysed and their corresponding
  integrated luminosities during the LEP2 period.}
\label{tab:lep2}
\end{table}

To evaluate the signal efficiencies and background contamination,
events were ge\-ne\-ra\-ted using different programs, all
relying on {\tt JETSET} 7.4 \cite{JETSET}, tuned
to LEP1 data \cite{TUNE} for quark fragmentation.

Slepton pair samples at 208 GeV centre-of-mass energy were
 produced with {\tt PYTHIA} 5.7\cite{JETSET} with
sleptons having mean decay lengths from  0.25 to 200 cm and masses
from 60 to 104\GeVcc. 
Other samples of slepton pairs were produced at 206 and 208 GeV with 
{\tt SUSYGEN}~\cite{SUSYGEN} for the small impact parameter search with 
 $m_{\tilde{\tau}}$\ from 90\GeVcc~to 102\GeVcc~and 
 $m_{\tilde{\mu}}$\ equal to 90\GeVcc. A sample of selectrons with mass equal
 to 90\GeVcc~was produced at 206 GeV.
Neutralino pair events and their subsequent decay products were generated 
with {\tt SUSYGEN}. Selection efficiencies were computed from samples with 
neutralino masses from 72\GeVcc~$\leq~m_{\tilde{l}}$ + 
2\GeVcc~$\le m_{\tilde{\chi}^\circ_1}~\le~\sqrt{s}$/2 at 206 GeV. 
{\tt SUSYGEN} was also used to generate the chargino pair production and
decay. In order to compute detection efficiencies, samples at 204~GeV and
206~GeV centre-of-mass energies were generated 
with gravitino masses of 1, 100 and 1000\eVcc ,
$m_{\tilde{l}}+ 0.3 \GeVcc \leq m_{\tilde{\chi}^+_1} \leq 
\sqrt{s}/2$\ and
 $80\GeVcc \leq m_{\tilde{l}}\leq \sqrt{s}/2-2.6\GeVcc$. Samples with
 smaller $\Delta m = m_{\tilde{\chi}^+_1} -m_{\tilde{\tau}_1}$\ were not
 generated because in that region each chargino decays 
into a W and a gravitino with an appreciable
branching ratio.

In the search for heavy stable charged particles, signal efficiencies were 
estimated from pair produced heavy smuons generated 
at energies of 205~GeV, 206.7~GeV and 208~GeV with {\tt
  SUSYGEN}. 
The events were passed through the detector simulation as heavy muons. The
efficiencies were estimated for masses between 10\GeVcc~and
100\GeVcc.  

For the sgoldstino search, signal efficiencies for the channel
$\phi~\rightarrow~\gamma \gamma$ were estimated from 
QED background events generated according to~\cite{sgoldstinoeffi}.
On the other hand, the selection efficiency for the $\phi~\rightarrow~g g$
channel was evaluated using $q\overline{q}\gamma$ background events generated
with {\tt PYTHIA} and processed through the full DELPHI analysis chain and
re-weighted according to the background and signal photon polar angle
distribution. 

The background process \eeto\qqbar ($n\gamma$) was generated with
{\tt PYTHIA 6.125}, while {\tt KORALZ 4.2} \cite{KORALZ} was used
for $\mu^+\mu^-(\gamma)$ and $\tau^+\tau^-(\gamma)$.
The generator {\tt BHWIDE}~\cite{BHWIDE} was used for \eeto\ee\ events.
Processes leading to four-fermion final states
were generated using {\tt EXCALIBUR 1.08}~\cite{EXCALIBUR} and {\tt
  GRC4F}~\cite{GRC4F}. 
Two-photon interactions leading to hadronic final states
were generated using {\tt TWOGAM}~\cite{TWOGAM}, including the VDM, QPM and
QCD components.
The generators
of Berends, Daverveldt and Kleiss~\cite{BDK} were used for the leptonic
final states.

The cosmic radiation background was studied using cosmic muons collected
before the beginning of the 2000 LEP run.

The generated signal and background events were passed through the
detailed simulation~\cite{performance}
of the DELPHI detector 
and then processed
with the same reconstruction and analysis programs used for real 
data.
%
%
\section{Data selection}
\label{dataselection}
The following sections describe the selection criteria used to search for the
different topologies summarized in Table~\ref{t:gmsbtop}. The searches for
slepton, neutralino and chargino pair production are based on the ones
described in~\cite{nuestro_papel_202,nuestro_papel_189,nuestro_papel_183,nuestro_papel_172,nuestro_papel_chargi_183}.
The search for heavy stable charged sleptons is based on the analysis already
published
in~\cite{nuestro_papel_202,heavyparticles,heavy_stable_120-183}. Finally the
sgoldstino search has already been presented in~\cite{ref:delphisgold}.

\subsection{Slepton pair production}

This section describes the selection criteria used in the search for the
process 
$e^+e^- \rightarrow \tilde{l}^+ \tilde{l}^- \rightarrow l^+ \tilde{G} l^-
\tilde{G}$. 
Loose preselection cuts were imposed on the events in order to 
suppress as much as possible the low energy background
(beam-gas and beam-wall) and the SM processes. 
The reconstructed tracks of charged  particles were 
required to satisfy certain quality criteria:
momenta above 100~MeV/c and impact parameters (geometric signed) 
below 4~cm in the plane transverse to the beam pipe ($xy$), and below 10~cm
in the direction along the beam pipe ($z$). 
Clusters in the calorimeters were interpreted as neutral
particles if they were not associated to charged particles and if their
energy exceeded 100~MeV. Only these particles were used to compute the
general event quantities. The preselection cuts were the following: 
\begin{itemize}
\item to eliminate high multiplicity events like \Zqq ($n\gamma$) or 
WW, W$e\nu_e$~and $ZZ$ when the produced particles had
pure hadronic or semileptonic decays, the multiplicity computed with the
particles that satisfied the quality requirements  was
required to be between 1 and 6 (the multiplicity of all signal samples was
very well contained between these two limits);
\item to eliminate two-photon processes, the visible energy in the event was 
required to be above 10 GeV; 
\item to eliminate the remaining contribution of two-photon and two-fermion 
processes, the absolute value of the transverse momentum vector of charged 
and neutral particles was required
to be  greater than 5\GeVc;
\item the energy measured in the very forward calorimeters (STIC)
 was required to be below 10 GeV to eliminate the residual contamination of
 processes mentioned above;
\item to eliminate the Bhabha contribution, the total electromagnetic
    energy was required to be less than the beam energy.
\end{itemize}
All the events that survive the preselection cuts underwent the search for
secondary vertices or kinks. Only the events which were not tagged as
kink candidates passed the selection criteria to search for large impact
parameter tracks.

\subsubsection{Search for secondary vertices or kinks}
\label{kink}
\hspace{\parindent}
The analysis exploits a peculiarity of the $\tilde{l}^\pm \rightarrow l^\pm
\tilde{G}$  topology in the case of intermediate gravitino masses 
(i.e. few\eVcc $< m_{\tilde{G}} <$ few hundred\eVcc), namely, one or
two tracks coming from the interaction point and at least one of them with
either a secondary vertex or a kink. 

All the charged particles 
of the event that survived the preselection cuts were
grouped into clusters (in order to group all the particles coming from a
tau decay) according to their first measured point in the $xy$ plane. This
clustering procedure was iterative and worked 
as follows. The pair of particles with the smallest separation at their 
respective starting points was considered first. If this separation was 
smaller than 2 cm, the particles were grouped to form a cluster whose starting 
point was defined as the average of their first measured points. The two 
particles were then replaced by this cluster which was subsequently  treated
as  
a pseudo-particle. The process was then repeated until all charged particles 
 or pseudo-particles were grouped into clusters. 
This procedure allowed for clusters containing
a single particle if its momentum was larger than 1.5\GeVc. 
Events were rejected if more than 6 particles were not grouped into clusters or
if a cluster could not be obtained. 
This cut was intended to eliminate the remaining beam 
related background events\footnote{This kind of event was mainly
  characterized by a high number of low momentum charged particles seen only
  in the innermost detectors VD and ID.} that had not been excluded at the
  preselection level.  

Once all the particles were grouped in clusters,
the search for kinks was performed in the following way. 
Slepton candidates were searched for among all the clusters in the
event. Among the remaining clusters,
the ones corresponding to lepton candidates or decay products of taus 
were also searched for. The clusters were ex\-tra\-po\-la\-ted in
order to find a crossing point. If the crossing point existed, the event was
considered as a kink candidate. 
Reconstruction of secondary vertices for the case $\tilde{\tau}\rightarrow
\tau \tilde{G}$ is illustrated
in Figure~\ref{fig.fig-grav-def}, which shows a decay vertex and
the variables used in the analysis.

Isolated particles (clusters with only one particle) were considered as 
\slep~candidates if their trajectories were compatible with particles coming from the 
interaction point according to the following selection criteria: 
\begin{itemize}
\item the first measured point with respect to the beam spot in the
    plane transverse to the beam axis
    ($R_{sp}^{\tilde{l}}$) had to lie in the Vertex Detector (VD);
\item the momentum of the particle was greater than 2\GeVc;
\item the polar angle with respect to the beam axis
had to satisfy $|\cos\theta|<0.8$, co\-rres\-pon\-ding to the barrel region;
\item the impact parameter 
along the beam axis and in
    the plane per\-pen\-di\-cu\-lar 
    to it was less than 10 and 4~cm, respectively. 
\end{itemize}
For every \slep~candidate, a search was
made for a second cluster satisfying the following selection criteria: 
\begin{itemize}
\item the starting point in the transverse plane 
($R_{sp}^{l_d}$) had to be greater than  $R_{sp}^{\tilde{l}}$.
The second cluster starting point 
was always found in the Inner Detector (ID) or 
the Time Projection Chamber (TPC); 
\item the angular separation between the directions defined by the
\slep~candidate and the lepton candidate had to be smaller than
45$^\circ$ in the {\it xy} plane, to consider only
 the particles which were in the direction of the slepton. 
\end{itemize}

The \slep~candidate and the lepton cluster had to define a common crossing
point, called secondary vertex or kink. If the lepton
cluster included more than one charged particle (which is the case when $l$
is a $\tau$ decaying to 3 or 5 prongs), only the one with the highest 
momentum was used to search for the kink. 
To find the crossing point, 
the particle trajectories were represented by a helix
in space. Taking into account this parametrization, the 
point of closest approach between the $\tilde{l}$ particle
and the selected particle from the $l_d$\ cluster was calculated.
The conditions to define a good crossing point between both 
particles were the following: 
\begin{itemize}
\item the minimum distance between the particles had to be smaller
        than 1~mm in the {\it xy} plane;
\item the crossing point, the end point of the slepton and the
        starting point of the lepton were required to satisfy
        the following conditions:
        \begin{eqnarray}
    -10 \,{\mathrm cm}\,\,&<\,\,(R_{cross}-R_{end}^{\tilde{l}})\,\,<&
        \,\,25\, {\mathrm cm} 
        \nonumber \\
    -25 \,{\mathrm cm}\,\,&<\,\,(R_{cross}-R_{sp}^{l})\,\,<& \,\,10\,
        {\mathrm cm}, 
        \end{eqnarray}
        \label{eq:cut2}
\noindent
        where $R_{end}^{\tilde{l}}$, $R_{cross}$ and $R_{sp}^{l}$
        are the distances w.r.t the beam spot
        of the end point of the slepton,
        the crossing point  and the starting point of the
        lepton in the {\it xy}  plane. The cut was optimised to
        assure that all VD or IDVD only particles (particles which only had ID
        or VD  hits) were connected with particles reconstructed with the
        TPC.   
\end{itemize}
\noindent
The resolution achieved (generated distance minus
reconstructed distance) with the algorithm to find secondary vertices in
 the coordinates $x$ and $y$ was 0.14 mm. The $z$ coordinate was not taken
 into account in the search for a crossing point because the resolution for
 low momentum particles is very poor.

Fake decay vertices could be present among the reconstructed secondary 
vertices. They could be produced by particles interacting in 
the detector material or by radiated photons when the particle trajectory
was reconstructed as two se\-pa\-ra\-ted particles.
To eliminate these events, additional conditions were
required:
\begin{itemize}
\item to reject hadronic interactions, the angle between the direction
        of any reconstructed hadronic vertex w.r.t. beam spot (secondary
        vertices  
        reconstructed in region where there is material) and the direction
        of the slepton candidate 
        must be greater than 5$^\circ$; 
\item to reject segmented tracks,
        the angle between the particles used to define a vertex
        had to be larger than 6$^\circ$; 
\item to reject photon radiation
        in the case of $l$ clusters with only one particle,
        there had to be no neutral particle  in a 3$^\circ$ cone
        around the direction defined by the difference between the
        $\tilde{l}$ momentum and the momentum of the
        $l$ calculated at the crossing point.
\end{itemize}
If no pair of particles was found to survive these conditions, the event was
rejected. Figure~\ref{fig:grav:kinks_BG} shows the distribution of these three
quantities. The distributions compare real data, expected SM background
simulation and a simulated signal for \mstau = 60\GeVcc~with a mean decay
length of 50~cm.  

Table~\ref{tab:kinkback} shows the different SM background
contributions and the observed events in data after applying the selection 
criteria to search for kinks. 
Efficiencies for different gravitino and stau masses were calculated by
applying the above selections to the simulated signal
samples. Figure~\ref{fig:effivsradio} shows  
the secondary vertex reconstruction efficiency as a function of the stau decay 
radius. For smuons and selectrons the same dependence is observed. For smuons
the efficiency plateau is around 60\%, while for selectrons it is around
40\% due to the preselection cut on total electromagnetic energy.  
\begin{table}[hbt]
\begin{center}
\begin{tabular}{||c|c||} \hline 
Observed events
             & 2                                            \\ \hline
Total background
     & 0.88$^{+1.35}_{-0.15}$                             \\ \hline    
\hline
$Z^*/\gamma \to (\tau \tau) (n\gamma)$
     & 0.16$^{+0.09}_{-0.05}$ \\ \hline
$Z^*/\gamma \to (ee) (n\gamma)$
   & 0.00$^{+0.87}_{-0.00}$ \\ \hline
4-fermion (except $\gamma\gamma$)
   & 0.12$^{+0.06}_{-0.01}$\\ \hline
$\gamma\gamma \to \tau^+\tau^-$
    & 0.16$^{+0.32}_{-0.06}$\\ \hline
$\gamma\gamma \to e^+e^-$
    & 0.44$^{+0.98}_{-0.11}$ \\ \hline
\end{tabular}
\end{center}
\caption[.]{
Number of observed events at $\sqrt{s}$ from 204~GeV to 208~GeV
together with the total number of expected SM background events
and the expected numbers from the individual background sources,
for the secondary vertex search. The asymmetric errors are due to the
Poissonian description of the statistics.}
\label{tab:kinkback}
\end{table}
    
\subsubsection{Large impact parameter search}
\label{largeip}
\hspace{\parindent}
To investigate the region of low gravitino masses (short decay lengths) the
previous search was extended to the case of sleptons decaying
between 0.25~cm and around 10~cm, i.e., before the tracking devices. 
In this case it was only possible to reconstruct the slepton decay products. 
The impact parameter search was only applied to those events accepted by the
same preselection cuts as in the search for secondary vertices but not 
selected by the vertex analysis. The events used
 in this search contained exactly two single particle clusters (i.e. two
 charged 
 particles with momentum larger than 1.5\GeVc~and a distance between starting
 points greater than 2 cm) which were acollinear and had large impact
 parameters. The events were accepted as candidates if: 
\begin{itemize}
\item the first measured point in the $xy$ plane of at least one of the
  particles was in the VD; 
\item both particles were reconstructed with the TPC to guarantee  good
       particle re\-cons\-truc\-tion quality;
\item at least one of the particles had an impact parameter w.r.t. the beam
  spot larger than 0.2~cm in the $xy$ plane to remove SM events;
\item the ratio of the maximum impact parameter over the minimum impact 
       parameter in the $xy$ plane  and w.r.t. the beam
  spot was
       smaller than -1.5 or larger than -0.5, to reject cosmic rays since
       they are characterized by large impact parameters of the same value and
       opposite sign. 
\item the acollinearity\footnote{The acollinearity is defined as being
    180$^\circ$ minus the angle between the momentum vectors of both
    particles.}
 between the two particles was larger than 10$^\circ$ to
eliminate back-to-back events with badly reconstructed particles or 
interactions which always gave small acollinearities.
In addition, to reduce further the 
cosmic ray background, the acollinearity between the two particles was
required to be smaller than 175$^{\circ}$, since an off-time cosmic ray   
crossing from one TPC drift half to the other could be  reconstructed 
as two almost parallel particles. 
\end{itemize}
Figure~\ref{fig:acoll} shows the acollinearity distribution for real data
minus simulated SM background, compared to cosmic rays for events which passed the preselection cuts and cuts 1-3. The latter follows
pretty well the acollinearity distribution of the former difference.

Table~\ref{tab:ipback} shows the different SM background contributions and the
number of events observed in data after applying the selection criteria to
search for large impact parameter particles.  
\begin{table}[hbt]
\begin{center}
\begin{tabular}{||c|c||} \hline 
Observed events
             & 2                                            \\ \hline
Total background
     & 2.40$^{+1.44}_{-0.36}$                             \\ \hline    
\hline
$Z^*/\gamma \to (\tau \tau) (n\gamma)$
     & 0.05$^{+0.05}_{-0.01}$ \\ \hline
$Z^*/\gamma \to (ee) (n\gamma)$
   & 0.12$^{+0.90}_{-0.10}$ \\ \hline
4-fermion (except $\gamma\gamma$)
   & 0.65$^{+0.09}_{-0.05}$\\ \hline
$\gamma\gamma \to \tau^+\tau^-$
    & 0.09$^{+0.31}_{-0.04}$\\ \hline
$\gamma\gamma \to e^+e^-$
    & 1.49$^{+1.08}_{-0.34}$ \\ \hline
\end{tabular}
\end{center}
\caption[.]{
Number of observed events at $\sqrt{s}$ from 204~GeV to 208 GeV
together with the total number of expected SM background events
and the expected numbers from the individual background sources,
for the large impact parameter track search. The asymmetric errors
are due to the Poissonian description of the statistics.}
\label{tab:ipback}
\end{table}
 
The efficiencies were derived for different slepton masses and decay lengths
by applying the same selection criteria to the simulated signal events. 
In the search for \staone~the ma\-xi\-mum efficiency was around 32\% corresponding 
to a mean decay  
length of 2.5~cm. The efficiency decreased sharply for lower decay lengths
due to the requirement on mi\-ni\-mum impact parameter. 
For longer decay lengths, 
the appearance of reconstructed $\tilde{l}$ in combination with
the cut on the maximum number of charged particles in the event caused the
efficiency to decrease smoothly. This decrease was compensated by a rising
efficiency in the search for secondary vertices.
For masses above 60\GeVcc\ no dependence on the  $\tilde{l}$ mass was 
found far from the kinematic limit.  

The same selection was applied to smuons and selectrons. For smuons
the efficiency increased to $\sim$ 58\% for a mean decay 
length of 2.5 cm and masses over 
60\GeVcc\ since the smuon always had a one-prong decay.
For selectrons the efficiency was $\sim$ 33\% for the same mean decay
length and range of masses. 

\subsubsection{Small impact parameter search}
\label{smallip}
\hspace{\parindent}

The large impact parameter search can be extended further to 
mean decay lengths below 0.1 cm. 
Charged particles were selected if their impact parameter was less than 10~cm 
in the plane transverse to the beam direction and less than 15~cm in the 
direction along the beam pipe. The polar angle had to be between 20$^\circ$ 
and 160$^\circ$. Their measured momentum was required to be larger than 
400\MeVc~with relative error less than 100\% and track length larger than 
30~cm. Any calorimetric deposit associated to a discarded charged particle 
was assumed to come from a neutral particle.

The search was restricted to events with 2 to 4 charged particles and missing 
energy larger than 0.3$\sqrt{s}$. The $\gamma\gamma$ events were suppressed by 
requiring a visible energy greater than 0.08$\sqrt{s}$ and a transverse missing
momentum greater than 0.03$\sqrt{s}$. The polar angle of the missing momentum 
was required to be between 30$^\circ$ and 150$^\circ$, and the total energy
in a cone of $30^{\circ}$, $E_{30}$, around the beam-pipe
was required to be less than 10\% of the visible energy, and the neutral
energy was required to  be less than 0.175$\sqrt{s}$.

The events were then divided into two hemispheres using the thrust
axis.
The 
total momentum of charged and neutral particles in each hemisphere was 
computed and used to define the acollinearity of the event. Standard 
{\mbox{$e^+e^-~\rightarrow\,f\overline{f}(\gamma)$}} processes and cosmic
rays were reduced by requiring the acollinearity to be greater than
 10$^\circ$. The charged particle with the largest and good quality momentum  
($\Delta p_i/p_i <$ 50\%) in each hemisphere was selected as the 
leading particle.
The following quality requirements were only applied to the leading particles: 
the first measured point of the particle tracks 
had to be within 50 cm of the beam spot  
in the $xy$ plane, the particles were required to have at least one segment 
beyond the ID detector and to be away from insensitive regions of the 
electromagnetic calorimeter. In addition, at least one of the leading
 particles was required to be reconstructed with the TPC.

{\mbox{$e^+e^-~\rightarrow\,f\overline{f}(\gamma)$}} processes and cosmic
rays 
were further reduced by requiring an angle between the leading particles in
the $xy$ plane of less than 3 radians. Hadronic events, in particular 
$\gamma\gamma~\rightarrow~q\overline{q}$ or, in general, events where the
available energy is shared by many particles (including
undetected/unselected), were rejected by requiring $\sqrt{p^2_1 + p^2_2} >$ 
0.03$\sqrt{s}$, where $p_1$ and $p_2$ are the momenta of the leading
particles.  
To reduce Bhabha events the total electromagnetic energy of the leading 
particles, $E_1+E_2$, had to be less than 0.35$\sqrt{s}$. By requiring that 
any leading particle with an impact parameter larger than 1 cm in the $xy$
plane and measured w.r.t. the beam spot 
should be reconstructed by the TPC and at least one other detector, the
residual cosmic rays (in particular the out-of-time cosmic rays) 
 were rejected. Finally, photon conversion events with only two
particles   
were rejected by requiring the angle between them at their perigee to be 
greater than 5$^\circ$. 

The background left after the selection described above consisted mainly of 
events containing $\tau$ pairs in the final state 
($\gamma^*/Z^* \rightarrow \tau\tau$ and WW $\rightarrow \tau\nu\tau\nu$). 
To reject these events, the variable
$\sqrt{b_1^2 + b_2^2}$, where  b$_1$ and b$_2$ are the impact parameters
in $xy$ (measured w.r.t. the beam spot) of 
the two leading particles, was used. Requiring $\sqrt{b_1^2 + b_2^2}
\geq$ 0.06~cm  eliminated most of the remaining background.

In order to preserve the efficiency in the region of decay length above
10~cm, where the \slep~can be observed as a particle coming from the 
primary vertex and badly measured due to its limited length, further 
requirements on the particle quality were applied only to the leading
particle with  
the largest impact parameter (measured w.r.t. the beam spot). 
This particle was required to have a relative 
momentum error $< 30\%$ and the particle had to be measured at least either in 
the TPC or in all of the other three tracking
detectors in the barrel (VD, ID and OD).

The efficiency of the search did not show any significant dependence 
on the \slep~mass for masses over 40 GeV/$c^2$ and far from the 
kinematic limit, and it could be parameterized as a function of  
the \slep~decay length in the laboratory system. 
The efficiency for \staone~detection
reaches $\sim$ 40\% for decay lengths around 2 cm. It is still 16\% for a
decay length of 0.1~cm, and 13\% for 20~cm. The efficiency for 
${\tilde{\mu}}$ detection reaches 45\% around 2~cm, 15\% at 0.1~cm, and 
17\% at 20~cm. 


In order to increase the efficiency in the search for selectrons, the cut 
\mbox{$(E_1+E_2)<0.35\sqrt{s}$}\ was not applied. The Bhabha events that 
survived the selection were those where at least one of the electrons 
un\-der\-went a secondary interaction, thus acquiring
a large impact parameter. However, it was found that in these cases the
measured momentum of the electron was smaller than the electromagnetic energy
deposition around the electron track. Therefore, the cut 
 $(E_1/p_1+E_2/p_2)<2.2$ was used for the selectron search.
The maximum efficiency reached in the 
selectron search was $\sim$~35\% at  $\sim$ 2~cm mean decay length.

The number of events selected in the data was 4 in the $\tilde{\tau}$ and
$\tilde{\mu}$ search. The same 4 events also passed the $\tilde{e}$ search. 
The expected SM background in both searches was 3.3$\pm$0.3 events. 
Figure~\ref{fig:ip-data-mc} shows the $\sqrt{b_1^2 + b_2^2}$ distribution for
data (dots), simulated SM backgrounds (grey histogram) and  
simulated signal of $m_{\tilde{\tau}_1}$ = 90 GeV/c$^2$ and
$m_{\tilde{G}}$ = 25\eVcc~at $\sqrt{s}$ = 206~GeV and
a boosted mean decay length of around 1~cm (white histogram)
after all other cuts.
This figure only shows two of the candidates. The other two events not shown
have  $\sqrt{b_1^2 + b_2^2}~\sim$ 1.5 cm. The overflow bin has 1.3
expected events from Monte Carlo and 2 from data. 
All the selected candidates were compatible with SM events. 

\subsection{Neutralino pair production}
\label{neutralinos}
\hspace{\parindent}
In this section, the selections used to search for the process
$e^+e^- \to \tilde{\chi}^0_1 \tilde{\chi}^0_1 \to 
\tilde{\tau}_1 \tau \tilde{\tau}_1 \tau  \to \tau \tilde{G} \tau \tau 
\tilde{G} \tau $\ within the \staone~NLSP scenario, and the process
$e^+e^- \to \tilde{\chi}^0_1 \tilde{\chi}^0_1 \to 
\tilde{l} l \tilde{l'} l'  \to l \tilde{G} l l' \tilde{G} l'$ (with
BR$(\tilde{\chi}^0_1 \to \tilde{l} l) = 1/3$\ for each leptonic flavour) 
within the co-NLSP scenario, are presented. 

In the following, the preselection of the events, common to both scenarios, 
is presented. 
The reconstructed tracks of charged particles were 
required to have momenta above 100\MeVc\ and impact parameters below 4~cm
in the transverse plane and below 10~cm in the lon\-gi\-tu\-di\-nal
direction. The relative error on the measurement of the momentum had to be
smaller than 100\%.
Clusters in the calorimeters were interpreted as neutral
particles if they were not associated to charged particles and if their
energy exceeded 100~MeV.
All charged and neutral particles that satisfy these criteria were 
considered good particles and they were used to compute the relevant event 
quantities.
To assure good quality of the data, the ratio of good to  
total number of particles was required to be above 0.7. 
Particles that did not pass quality selection but had an
associated calorimetric energy of at least 2\GeV~had their angles 
taken from those of the track, but their momentum
was recomputed from the energy of the calorimetric measurement 
(such particles were not included in the good sample).
Events had to have between four and ten good charged particle tracks.
In addition, it was required that the thrust be less than 0.99;
the transverse momentum
had to
be bigger than 3\GeVc, and $|\cos\theta_{p_{miss}}|<0.95$ (polar 
angle of the missing momentum vector).
Very forward-going events
were eliminated by requiring $E_{30}$ less than 
70\% of the total visible energy.
With this preselection, the total number of simulated background events and
real data events was reduced by a factor 
of about 6000. Only events passing these pre-selections were 
analysed further.

The selection takes advantage of the fact that 
signal events can be separated into two different kinematic regions of
the ($m_{\tilde{\chi}^0_1}$,$m_{\tilde{l}}$) space: when the 
mass difference $\Delta m =   m_{\tilde{\chi}^0_1} - m_{\tilde{l}}$\ 
is bigger than about 10\GeVcc, all four leptons carry similar momenta. 
When the difference becomes smaller, the two leptons coming from the 
decay of the $\tilde{l}$\ tend to be the most energetic, 
increasingly so as the ${\tilde{\chi}^0_1}$\ mass increases. The Durham 
algorithm~\cite{Durham} was used to 
divide the event into four jets
by allowing the jet resolution parameter
to vary as a free variable. Numbering the jets from 1 to 4 with 
${\rm E}_1 > {\rm E}_2 > {\rm E}_3 > {\rm E}_4$, a variable
$r$ was defined as:
\begin{equation}
r = \frac{{\rm E}_3 + {\rm E}_4}{{\rm E}_1 + {\rm E}_2} \ .
\end{equation}
The distribution of $r$\
shifts towards lower values with increasing neutralino masses.

At the preselection level the two main differences between the
\staone~NLSP and \slepr~co-NLSP scenarios
come from the fact that the mean number of neutrinos carrying away
undetected energy and momentum and the number of charged particles per event is
considerably bigger for the former scenario. 

In the \staone~NLSP scenario, the simulated background samples were then
divided into two samples above and below $r = 0.1$ and different requirements
were imposed in the two cases. No significant dependence on this
variable was observed in the \slepr~co-NLSP scenario.
Two sets of cuts were applied in order to reduce the 
$\gamma\gamma$\ and ${\rm f}{\bar{{\rm f}}}(\gamma)$\ backgrounds
and a third set of cuts to select events according to their topology.
Those cuts are compiled in Table~\ref{tab:neutralinocuts}.

\begin{table}[h]
\begin{center}
\begin{tabular}{|c|c|c|c|c|}\cline{3-5}
\multicolumn{2}{c}{} & \multicolumn{2}{|c|}{\staone~NLSP} & \sLepR~co-NLSP \\
\cline{3-4}
\multicolumn{2}{c|}{} & $r >$ 0.1 & $r~\leq$ 0.1 & \\ \hline
Cuts    & $E_T$ (GeV)    & $>$ 11 & $>$ 12 & $>$ 4 \\ 
against & $E_{30}$ (GeV) & $<$ 0.6$\sqrt{s}$ & $<$ 0.6$\sqrt{s}$ & $<$ 0.6$\sqrt{s}$ \\
$\gamma\gamma$\ & $m_{miss}$ (GeV) & $<$ 0.88$\sqrt{s}$ & $<$ 0.9$\sqrt{s}$ & $<$ 0.88$\sqrt{s}$ \\ 
        & $p_{max}$ (\GeVc) & $>$ 4 & $>$ 3 & $>$ 8 \\
        & $p_T$ (\GeVc) &  &  & $>$ 6 \\ \hline
Cuts    & $N_{good}$ & $<$ 7 & $<$ 9 & $<$ 7 \\
against & $T$ & $<$ 0.975 & $<$ 0.975 & $<$ 0.95 \\
${\rm f}{\bar{{\rm f}}}(\gamma)$ & acoplanarity ($^\circ$) & $>$ 8 & $>$ 8 & $>$ 8
\\
 & $m_{miss}$ (GeV) & $>$ 0.3$\sqrt{s}$ & $>$ 0.3$\sqrt{s}$ & $>$ 0.2$\sqrt{s}$ \\ \hline  
Cuts & {\it jet-beam angle} ($^\circ$) & $>$ 17 & $>$ 17 & $>$ 18 \\
based on & {\it 2-jets cone angle} ($^\circ$) & $>$ 20 & $>$ 20 & $>$ 25 \\
topology & {\it 4-jets separation} ($^\circ$) & $>$ 8 & $>$ 4 & $>$ 9 \\ \hline
\end{tabular}
\caption[]{Sets of cuts applied in the search for neutralino pair production
  to eliminate the different background sources. $p_{max}$ is the momentum of
  the charged particle with largest momentum; $N_{good}$ is the number of
  good particles in the event; $T$ is the event thrust; $m_{miss}$ is the 
missing mass of the 
event. Signal events tend naturally to cluster into a 4-jet topology. Taking
  this into account, cuts based on topology were applied: {\it jet-beam angle} is
  the angle between the jet direction and the beam direction; {\it 2-jets cone
  angle} represents the broadness of the 2-jets (considering only charged
  particles) obtained when reducing the
  4-jet topology into a 2-jets configuration with the Durham algorithm;
  and finally, {\it 4-jets separation} is the angle between jets.} 
\label{tab:neutralinocuts}
\end{center}
\end{table}

After these cuts, an efficiency between 26 and 44\% was obtained for 
the signal events in the \staone~NLSP scenario, and between 35 and 46\% in the
$\tilde{l}$ co-NLSP scenario. The number of 
events remaining in data and simulated samples after the 
selection procedure were 8 and 7.1$\pm$0.6 respectively in the \staone~NLSP scenario, and 7 and 6.6$\pm$0.6 respectively in the \slep~co-NLSP scenario.

\subsection{Chargino pair production}
\label{charginos}
\hspace{\parindent}
The search for chargino pair production, 
$e^+e^-~\rightarrow~\tilde{\chi}^+_1 \tilde{\chi}^-_1~\rightarrow~\tilde{l}^+
\nu \tilde{l}^- \overline{\nu}~\rightarrow~l^+ \nu \tilde{G} l^-
\overline{\nu} \tilde{G}$, makes use, without modification, of
four different analyses depending on the gravitino mass or, equivalently, on
the mean decay length of the slepton. When the slepton decays at the vertex,
the combination of two analyses can be exploited, the search for charginos and
the search for acoplanar leptons in gravity mediated supersymmetry breaking
scenarios. Details of 
these analyses can be found in~\cite{charg-msugra,slept-msugra}. For
intermediate mean decay lengths of the slepton the topology is large impact
parameter tracks or kinks, therefore, the two analyses explained in
sections \ref{kink} and \ref{largeip} can be used. Finally, if the slepton
decays outside the tracking devices the signature corresponds to stable heavy
leptons and this analysis is explained in section \ref{heavystable}.

The selections developed for these searches were thus applied to
simulated data samples with different gravitino masses, and 
the results are presented in terms of 95\% confidence level (CL)
excluded regions in the ($m_{\tilde{l}},m_{\tilde{\chi}_1^+}$) 
plane in section \ref{chargino-results}.
    
\subsection{Heavy stable charged particles search}
\label{heavystable}
\hspace{\parindent}
In this analysis it is assumed that the slepton lifetime is large enough that
the sleptons can pass through the tracking devices without decaying. If the
sleptons are pair produced, one expects two slowly moving particles, which are
characterized  by an abnormally large 
energy loss in the TPC. Furthermore, no Cherenkov light is expected to be
produced by them in the 
Ring Imaging CHerenkov detectors (RICH) of DELPHI. Both set of information
are combined to identify such particles. 

Only events with two or three charged particles were considered. 
Candidate particles were required to have hits both in the VD and in the TPC.
The particle momentum had to be above 5~GeV/c$^2$, and the particle length
was required to be at least 30~cm. 
To reduce cosmic rays background the absolute value of the particle impact
parameter was required to be below 
0.15 cm in the $xy$ plane and 1.5 cm in $z$.
Furthermore, the difference of the impact parameters in $z$ of the two most energetic particles in the
event was required to be less than 1~cm. An additional protection was added against showering electrons for which the
particle extrapolation through the RICH can be unreliable. For charged particles with an associated electromagnetic
energy exceeding 15~GeV, an additional hit in the Outer detector was required.
A careful run selection ensured that the RICH detectors were fully operational. 

The energy loss measurement was required to be based on at least 80 wires, and
the dE/dx calibration was checked on a TPC sector by sector basis, using $Z^0$ calibration data of the same year. 

Events were selected if they contained at least one charged particle with:

\noindent
{\bf (I)} momentum above 5~GeV/c, high ionization loss in the TPC and no
  associated $\gamma$s in the gas radiator of the RICH (gas veto) or,  

\noindent
{\bf (II)} momentum above 15~GeV/c, ionization loss
 at least 0.3 below the expectation for a proton and surviving the gas veto
 or, 

\noindent
{\bf (III)} momentum above 15~GeV/c, surviving the gas
and the liquid RICH veto.

\noindent
An event was also selected if the two particles with the highest momentum
were characterized by a high ionization loss or a gas veto, or if both particles 
had a low ionization loss. Special care has been taken about the
dE/dx in sector 6 (S6) of the Time Projection Chamber which was not
operational during the second half of data taking in 2000, 
reducing the efficiency of the search by several percent as the dE/dx search
windows could not be applied to particles pointing to S6.
To recover some sensitivity for particles pointing to this sector, only the double
veto search window was applied, requiring hits in the Vertex Detector, Inner
Detector and Outer Detector to
ensure a good propagation of the particles through the RICH.

The data of the year 2000 has been subdivided into 3 energy bins, corresponding to energies below 206~GeV
(85 pb$^{-1}$), between 206~GeV and 207~GeV (124 pb$^{-1}$), and above 207~GeV (11.4~pb$^{-1}$). 
A total background of 0.25$\pm$0.04 events was estimated from data itself by
counting the number of particles passing the individual selection criteria.

No candidate events were selected in data. 
Figure~\ref{rare208} shows the data and the three main search windows.
The expectation for a 95~GeV/$c^2$ mass signal is also shown. Signal
efficiencies were estimated from simulation. 
For particle masses between 10\GeVcc~and 60\GeVcc~the signal efficiencies are
of the order 
of 30\%. For larger masses they rise with increasing mass to about 76-78\%.
Then the efficiency drops rapidly when approaching the kinematic limit,
and it is assumed to be zero at the kinematic limit.

\subsection{Sgoldstino search}
\hspace{\parindent}
This section describes the search for $e^+e^-~\rightarrow~\phi \gamma$
events, with the sgoldstino going to two gammas or two gluons. The two
channels considered here give rise to two different topologies. On the 
one hand,
if the sgoldstino decays into two photons, the final topology of the event is
three high energy photons, one of them monochromatic. On the other hand, if the
sgoldstino decays into two gluons, one monochromatic
photon and two jets can be expected in the final state.
 
\subsubsection{$\phi~\rightarrow~\gamma \gamma$ channel}
\hspace{\parindent}
Events were selected as $\gamma \gamma \gamma$ candidates if they 
had at least two electromagnetic energy clusters  
with $0.219<E / \sqrt{s}<0.713$; 
at least another one with $E>5 $ GeV and no more than 
two additional clusters, the second (if present) with $E<5 $ GeV.
The two most energetic electromagnetic clusters had to be in the HPC region, 
$42^{\circ}<\theta<89^{\circ}$ ($91^{\circ}<\theta<138^{\circ}$), or in the FEMC region,
$25^{\circ}<\theta<32.4^{\circ}$ ($147.6^{\circ}<\theta<155^{\circ}$). 
Finally, the third  cluster had to be in the region $\theta >$ 42$^{\circ}$ 
($\theta <$ 138$^{\circ}$) or $20^{\circ}<\theta<35^{\circ}$
($145^{\circ}<\theta<160^{\circ}$). 
The event should not have hits in two of the three  Vertex Detector layers
compatible (within $\pm 2^{\circ}$ in the azimuthal direction)
 with the extrapolated trajectory of a particle from the  
beam crossing point to an electromagnetic cluster in the calorimeters.

Further, two hemispheres were defined by a plane orthogonal to the direction
of the most energetic cluster. One hemisphere was
required to have no charged particles detected in the barrel region 
of the tracking  devices other than the VD with a momentum greater than
1\GeVc~extrapolating to within 5~cm of the mean beam crossing point.
The requirement was further strengthened to suppress
the larger $e^+e^-$ background, by demanding that
both hemispheres had no such particle  detected by the TPC with
$\theta<35^{\circ}$. 

The events obtained after this selection had a three-body final state
kinematics if there was no significant initial state radiation lost along the
beam pipe. 
Defining
$\Delta= \left| \delta_{12}\right| +\left| \delta_{13}\right| +\left|
  \delta_{23}\right| $, where $\delta_{ij}$
is the angle between the particles $i$ and $j$, 
$\Delta$ should be $360^{\circ}$ in a three-body final state since the
particles lie in a plane.  
If only the events with $\Delta>358^{\circ}$  were accepted,
the energies of the 
particles  could then be determined with very good precision on the basis of the
measured photon directions:
\begin{equation}
 E_1= \sqrt{s} \frac{\sin \delta_{23}}{\delta};~E_2= \sqrt{s} \frac{\sin \delta_{13}}{\delta};~
 E_3= \sqrt{s} \frac{\sin \delta_{12}}{\delta}
\label{ecorr}
\end{equation}
with $\delta = \sin \delta_{12}+ \sin \delta_{13}+  \sin \delta_{23}$. The error on the energy 
evaluation was further minimised by requiring 
min$(\delta_{12},\delta_{13},\delta_{23})>2^{\circ}$.

In $\phi \gamma$ events the $\phi$ decay products are expected to be
isotropically 
distributed in the $\phi$ centre-of-mass system. This fact implies that 
the distribution of $cos \alpha$, where $\alpha$ is the angle between the
$\phi$ direction (opposite to the prompt photon) and the direction of one of
the 
two  $\phi$ decay products, in the $\phi$ centre-of-mass system, should be
flat. On the other hand, in the QED 
background, $\left| cos \alpha \right|$  peaks at 1
and therefore only the combinations giving $\left | cos \alpha  \right |<0.9$
were accepted. 

The number of selected events giving up to three combinations and the expected
background were 22 and $20.3_{-1.9}^{+1.5}$, respectively.
The error on the background was due to a correction applied in order to take
into account the missing higher orders (additional radiation above $\alpha^3$
which gives events having low values of $\Delta$) in the simulation of the
QED background. These kind of events were removed only from the selected
sample of real data and therefore, a corresponding normalization correction
factor of 
$(-13^{+4}_{-7}) \%$ was applied to the simulated sample. This correction is
the dominant contribution to the systematic error.
No significant background in addition to $\eeggg$ events was found.

No significant variation in the  acceptance for a $\phi \gamma$ signal
and in the selection efficiency inside the acceptance region were
observed in the 2000 data with respect to the lower energies in previous 
years: acceptance $(51\pm 2) \%$ and efficiency $(76.6 \pm 2.5) \% $.

The energy resolution remained also unchanged
and it was better than 0.5$\%$ over the whole photon energy
range.

The  photon 
recoil mass spectrum obtained for  the events collected during the 2000
run including those reported in 
\cite{ref:delphisgold} is shown in Figure~\ref{fig:msggg}-a.
The data are superimposed on the expected QED background
distribution. 

\subsubsection{$\phi~\rightarrow~g g$ channel}

This channel is expected to give a final state with one photon 
and two jets. An event was selected as a $\gamma g g$ candidate 
if it had an electromagnetic energy  cluster identified as photon with 
$E> 5$ GeV and $\theta>20^{\circ}$. The event must not have
electromagnetic clusters below $\theta=5^{\circ}$. 
The total multiplicity (charged and neutral) had to be greater than $10$, and 
the charged multiplicity  greater than $ 5$.
To remove $\gamma \gamma$ events the cut 
$\sum_{i=1}^n \sqrt{(p_x^2+p_y^2)_i} >$ 0.125$\sqrt{s}$ (where $n$ is the
total multiplicity) was used.
The sum of the absolute values of all particle momenta 
along the thrust axis had to be greater than 0.20$\sqrt{s}$. 
An electromagnetic cluster with $E< 0.45\sqrt{s}$,
or a total particle multiplicity greater than 16 when the cluster energy was
greater than $0.45\sqrt{s}$ had to be present. 
The polar angle of the missing momentum had to satisfy 
$|cos(\theta_{p_{miss}})|~<$ 0.995. 
The visible energy had to be greater than 0.60$\sqrt{s}$. 
%
The jets had to be incompatible with the $b\bar{b}$ hypothesis by requiring 
the combined btag of the events to be less than zero\cite{btag}.
Finally, the aforementioned $|\cos{\alpha}|$ and  $\Delta$ had to
be less than 0.9 and greater than 350$^\circ$, respectively.     

The events were reconstructed forcing all particles but the photon 
into a 2-jet topology using the Durham~\cite{Durham} algorithm.
Events were removed if $ y_{cut}>0.02$.
The events were also rejected if the angle between the photon
and the nearest jet was less than $10^{\circ}$.
In the case of more than one photon candidate in the event, the 
most energetic one was considered as the one produced in 
$e^+e^-\rightarrow \phi \gamma$.

Similar to the $\gamma \gamma \gamma$ selection, 
the events obtained after 
this selection were three-body final state events 
in absence of additional lost radiation. 
Therefore  kinematic constraints
were applied here as well. 
In this case, however,
the jet direction was determined with a poorer precision  than that obtained
for photons, therefore the cut in $\Delta$ was less stringent and
the  resolution for the reconstructed photon energy was  poorer:
a two-Gaussian fit gave $\sigma_1 = 1.2$ GeV ($55 \%$ of the area) and
$\sigma_2 = 4.1$ GeV.  

The number of selected events 
and the expected background were 766 and $775 \pm 5$, respectively.

No significant variation in the  acceptance 
and in the selection efficiency inside the acceptance region was
observed in the 2000 data compared to the values at lower energies.
The acceptance was $(76\pm 2) \%$ (almost independent of $m_{\phi}$) 
and the efficiency ranged from 20 to 55$\%$
depending on the photon energy. 
The energy resolution was also unchanged.

The  photon recoil mass spectrum obtained for  the events collected during
the 2000 run and including those reported in
\cite{ref:delphisgold} is shown in Figure~\ref{fig:msggg}-b. 
The data are superimposed on the expected background distribution. 
\section{Results and interpretation}
\label{sec:resultados}

Since there was no evidence for a signal above the expected background, the
number of candidates in data and the expected number of background events
were used to set limits at the 95\% confidence level (CL) on the pair
production cross-section and masses of the sparticles searched for.
The model described in reference~\cite{Dutta} was
used to derive limits within the GMSB scenarios. This model  
assumes radiatively broken electroweak symmetry and 
null trilinear couplings at the messenger scale. The 
corresponding pa\-ra\-me\-ter space was scanned as follows:
$1\leq n \leq 4$, $5\ {\rm TeV}\leq\Lambda\leq 90\ {\rm TeV}$, 
$1.1\leq M/\Lambda \leq 10^9$, $1.1\leq \tan\beta\leq 50$, and 
$sign(\mu) = \pm$ 1. The meaning of the parameters is explained in
section~\ref{sec:intro}. 
 The limits presented here are at \mbox{$\sqrt{s}$ = 208 GeV}
after combining these results with those of the searches at lower
centre-of-mass energies using the likelihood ratio method~\cite{Read}.
\subsection{Slepton pair production}
\hspace{\parindent}
The results of the search for slepton pair production are presented in the 
($m_{\tilde{G}}$,$m_{\tilde{l}}$) plane in Figure~\ref{fig:excl_208}-a
combining the impact parameter, the kink and the stable heavy lepton
analyses, and using all DELPHI data from 130 GeV to 208 GeV
centre-of-mass energies~\cite{nuestro_papel_202,nuestro_papel_189,nuestro_papel_183,nuestro_papel_172}.  
 
The $\tilde{\tau}_1$ pair production cross-section depends on the mixing in
the stau sector. Therefore, in order to put limits on the $\tilde{\tau}_1$
mass, the mixing angle had to be fixed. The results presented here 
correspond to a mixing angle in the stau sector
which gives the minimum $\tilde{\tau}_1$ pair production cross-section.
Within the \stuno~NLSP scenario, the impact parameter and kink
analyses extended the
limit $m_{\tilde{\tau}_1} >$ 82.5\GeVcc~for {$m_{\tilde{G}} \lesssim$
  6\eVcc, set by MSUGRA searches~\cite{susana_paper}, up to
  \mbox{$m_{\tilde{G}}$ 
    = 600\eVcc}, reaching the maximum excluded value of
\mbox{$m_{\tilde{\tau}_1}$= 93.6\GeVcc}~for \mbox{$m_{\tilde{G}}$= 130\eVcc}. 
For \mbox{$m_{\tilde{G}} >$ 130\eVcc}~the best lower mass limit was set by the
stable heavy lepton search. 
Although not shown in the figures, DELPHI excludes $m_{\tilde{G}}$ from
3$\cdot$10$^{-4}$\eVcc\ to 0.2\eVcc. $m_{\tilde{\tau}_1}$ masses below 40\GeVcc\ have
  been already excluded as discussed in~\cite{nuestro_papel_189}.

Within the sleptons co-NLSP scenario, the cross-section limits were used to
derive lower limits for \slepr\ (Figure~\ref{fig:excl_208}-b) masses at 
95\%~CL. Assuming mass degeneracy between the sleptons, the kink and 
impact parameter searches 
extended the limit $m_{\tilde{l}_R} >$ 88\GeVcc~set by MSUGRA 
searches~\cite{susana_paper}
for very short NLSP lifetimes up to $m_{\tilde{G}}$= 800\eVcc. 
For the MSUGRA case the best limit from
the $\tilde{\mu}_R$ has been used. The maximum excluded value
of $m_{\tilde{l}_R}$= 96.5\GeVcc~was achieved for $m_{\tilde{G}}$=
150\eVcc. For 
$m_{\tilde{G}} >$ 150\eVcc~the best lower mass limit was set by the stable
heavy  lepton search. 
$\tilde{l}_R$\ masses below 40\GeVcc\ were excluded by LEP1
data~\cite{lep1ex}. 
In the case of $\tilde{l}_R$\ degeneracy, this limit improved to 43\GeVcc.

\subsection{Neutralino pair production}
\hspace{\parindent}

Limits for neutralino pair production cross-section were derived in 
the \stuno~NLSP and sleptons co-NLSP scenarios for
each ($m_{\tilde{\chi}_1^0}$,$m_{\tilde{l}}$) combination for $m_{\tilde{G}}$
below a few \eVcc. For the 
\stuno~NLSP case the combination took into account 
the results from the LEP runs from 1996 (for \mbox{$\sqrt{s}~\ge$ 161 GeV}) to
2000~\cite{nuestro_papel_202,nuestro_papel_189,nuestro_papel_183}.  
The limits for the production cross-section allowed 
some sectors of the ($m_{\tilde{\chi}^0_1},m_{\tilde{l}}$) space to be
excluded. In order to exclude as much as possible of the mass plane, the
results from two other analyses were taken into account. 
The first is the search for slepton pair production in the context of MSUGRA 
models.  
From the experimental lower limit on
the mass of the \staone~set by that search, $m_{\tilde{\tau}_1} >$
82.5\GeVcc~\cite{susana_paper}, it can be concluded that at least neutralino
masses below the lower bound for $m_{\tilde{\tau}_1}$ plus the tau mass are
excluded.  
The second is the search for lightest
neutralino pair production in the region of the mass space where  
$\tilde{\chi}_1^0$\ is the NLSP~\cite{2gamma_189} 
(the region above the diagonal 
line in Figure~\ref{fig:masses}, i.e. $m_{\tilde{\tau}_1} >
m_{\tilde{\chi}^0_1}$).  
Within this zone, the neutralino decays into a gravitino and a photon. 

As an illustration, Figure~\ref{fig:masses}
presents the 95\% CL excluded areas for $m_{\tilde{G}} <$ 1\eVcc~in the 
$m_{\tilde{\chi}_1^0}$ vs $m_{\tilde{\tau}_1}$ plane for the \stuno~NLSP 
scenario and for 
different values of the number of messenger generations ($n$).
The positive-slope dashed area is excluded by the neutralino pair production
search. 
The negative-slope dashed area is excluded by
the analysis searching for neutralino pair production followed by the decay
$\tilde{\chi}^0_1\rightarrow \tilde{G}\gamma$. 
The point-hatched area is excluded by the direct search for slepton pair
production within MSUGRA scenarios.
The neutralino mass range is plotted up to almost the kinematic limit.

\subsection{Chargino pair production}
\label{chargino-results}
\hspace{\parindent}

Limits on the production cross-section for chargino pairs were derived for
each ($m_{\tilde{G}},m_{\tilde{l}},m_{\tilde{\chi}_1^+}$) combination for
\mgrav~below a few \eVcc. 
Figure~\ref{fig:xsec-masss-chargi} (a)
shows, as an example, the 95\% CL upper limit on the chargino pair 
production cross-section at $\sqrt{s} = 208$~GeV as a
function of $m_{\tilde{\chi}_1^+}$\ and $m_{\tilde{l}_R}$\ 
after combining with the results of the searches for large impact parameter and
kink at lower energies using the likelihood ratio method~\cite{Read},
for $m_{\tilde{G}}$ = 100\eVcc. 
The limits on the chargino pair production cross-section 
were used to exclude areas within the 
($m_{\tilde{\chi}_1^+},m_{\tilde{l}}$) plane for different 
domains of the gravitino mass, combining results from 
all the centre-of-mass energies from 183~GeV to 208~GeV~\cite{nuestro_papel_202,nuestro_papel_189,nuestro_papel_chargi_183}.
Figure~\ref{fig:xsec-masss-chargi} also shows the regions excluded at 
95\% CL in the 
($m_{\tilde{\chi}_1^+}$,$m_{\tilde{\tau}_1}$) plane, 
for the \staone~NLSP scenario (b), and in the
($m_{\tilde{\chi}_1^+}$,$m_{\tilde{l}_R}$) plane, 
for the \slepr~co-NLSP scenario (c).
The positive-slope area is excluded for $m_{\tilde{G}}~>$ 1\eVcc.
This limit is also valid for 
smaller masses of the gravitino because they lead to the same final state 
topologies. The negative-slope area is only excluded 
\mbox{for $m_{\tilde{G}} >$ 100\eVcc}.
The areas below $m_{\tilde{\tau}_1}$=~82.5\GeVcc\ in the \staone~NLSP
scenario, and below $m_{\tilde{l}_R}$=~88\GeVcc\ in the \slepr~co-NLSP 
scenario, are excluded by the direct search for slepton pair production in
MSUGRA models~\cite{susana_paper}. 
These lower bounds on $m_{\tilde{\tau}_1}$ and $m_{\tilde{l}_R}$ from MSUGRA 
mean a limit on the $m_{\tilde{\chi}_1^\pm}$ at 82.5\GeVcc\ and 88\GeVcc,
respectively. In Figure~\ref{fig:xsec-masss-chargi} the exclusion regions
given by the search for charginos in GMSB scenarios start at those values.
The area of $\Delta m  \leq  0.3$\GeVcc\ is not excluded because 
in this region the charginos do not decay mainly 
to \stuno\ and $\nu_{\tau}$, but to W and $\tilde{G}$.
Thus, if 
$\Delta m\! \geq\! 0.3$\GeVcc, the chargino mass   
limits are 100\GeVcc\ for all $m_{\tilde{G}}$, and 102\GeVcc\ for
$m_{\tilde{G}}~>$ 100\eVcc, in the ${\tilde{\tau}_1}$ NLSP
scenario. 
In the sleptons co-NLSP scenario the limits are 96\GeVcc~for
\mbox{$m_{\tilde{G}}=$ 1\eVcc}, and 102\GeVcc~for $m_{\tilde{G}} >$ 100\eVcc.
The chargino mass limit 
decreases with decreasing $m_{\tilde{\tau}_1}$\   because 
in scenarios with gravitino LSP small stau masses correspond to 
small sneutrino masses (both are proportional to $\Lambda$) 
and hence to smaller production 
cross-sections due to the destructive interference between the 
$s$- and $t$-channels.

It should be noticed that within the parameter space studied here, 
the lightest chargino is at least 40\% heavier than the lightest
neutralino. 
Thus, for small gravitino masses the search for neutralinos implies a
lower 
limit
on the lightest chargino of 130\GeVcc. Neutralinos are not directly
searched 
for in heavier gravitino mass regions and therefore for this range of
gravitino masses the limit of 102\GeVcc\ for the chargino mass remains valid.
\subsection{Heavy stable charged particle pair production} 
\hspace{\parindent}
The results presented in section~\ref{heavystable} were combined with 
previous DELPHI results in this
channel~\cite{nuestro_papel_202,heavyparticles,heavy_stable_120-183}, 
and cross-section limits were derived within MSSM models
as indicated in Figure~\ref{limit}. From the intersection points with
the predicted cross-sections for smuon or staus in the MSSM, left(right)
handed smuons and staus can be excluded up to masses
of 97.7(97.4)~GeV/$c^2$ at 95\%CL. No limits are given on selectrons here
because the cross-section can be highly suppressed by an additional
t-channel sneutrino exchange contribution.

\subsection{Sgoldstino production} 
\hspace{\parindent}
No excess of events nor clear evidence of an  
anomalous production of events with monochromatic photons is observed in 
either of the two channels.
Therefore   a limit on the cross-section of the new physics reaction
contributing to the two topologies was set.

The number of detected events, the background rate 
and the detection efficiency depend
on the $\phi$ mass. In addition, when the expected total
width for the same $m_\phi$ value is comparable or  larger than the 
experimental resolution, the data were compared with the background events
in a region corresponding to $80 \%$ of the signal area. 
As a consequence, the limit on the signal cross-section depends 
on $m_{\phi}$ and $\sqrt{F}$. 
Furthermore, to take into account  the different sensitivities of the two
analysed channels, the likelihood ratio method 
was used~\cite{Read}. 
Since the expected $\phi$  branching ratio 
and total width depend on the mass 
parameters as explained above, the 
$95\%$ CL  cross-section limit was computed as
a function of $m_{\phi}$ and $\sqrt{F}$ for the two sets of parameters listed 
in Table~\ref{tab:param}, and is shown in Figure~\ref{cslim}. 
By comparing the experimental limits with the expected production
cross-section, it is possible to determine a  $95 \%$
CL excluded region on the parameter space as shown in
Figure~\ref{excl}. As explained in~\cite{ref:prz},
to keep the particle interpretation the total width $\Gamma$ must be 
much smaller than $m_{\phi}$ and therefore the region with 
$\Gamma > 0.5 m_{\phi}$ was not considered.
In Figures~\ref{cslim} and~\ref{excl} the sgoldstino mass range starts at
10\GeVcc~to avoid all theoretical subtleties connected with the
non-perturbative aspects of the strong interaction. The upper value of the
mass range approaches the kinematic Al limit.  
\subsection{Limits on the GMSB parameter space} 
\hspace{\parindent}

Finally, all these results can be combined to produce exclusion plots 
within the ($\tan\beta, \Lambda$) space. 
The corresponding pa\-ra\-me\-ter space was scanned as follows:
$1\leq n \leq 4$, $5\ {\rm TeV}\leq\Lambda\leq 90\ {\rm TeV}$, 
$1.1\leq M/\Lambda \leq 10^9$, $1.1\leq \tan\beta\leq 50$, and 
$sign(\mu) = \pm 1 $.
As an example,  Figure~\ref{fig:lambda} shows the zones
 excluded for $n=$1 to 4 for $m_{\tilde{G}} \leq 1$\eVcc , which corresponds to
the NLSP decaying at the main vertex. 
The shaded areas are excluded. The areas below 
the dashed lines contain points of the GMSB parameter space with \nuno~NLSP.
The areas to the right (above for $n = 1$) of the dashed-dotted lines 
contain points of the GMSB parameter space where sleptons are the NLSP. 
It can be seen that the region of slepton NLSP increases with $n$. 
The contrary occurs to the region of neutralino NLSP. 

The shaded areas below the dashed lines are excluded by the search
for neutralino pair production followed by the decay
$\tilde{\chi}^0_1\rightarrow \tilde{G}\gamma$ 
(acoplanar photons search). However, the solid and almost vertical line
that sets the lowest limit to the parameter $\Lambda$, is defined by the
combination of the 
search for acoplanar photons, right handed selectrons with sleptons NLSP and
neutralino pair production with stau NLSP.  Finally, the search for
stau NLSP improves the exclusion limit represented by the solid line with
positive slope.

Theoretical arguments
constrain the value of $\tan\beta$ to be between $\sim$ 1.2 and $\sim$
65~\cite{martin}. A lower limit is set for the variable $\Lambda$\ at 17.5~TeV.

\section{Summary}
\hspace{\parindent}

Lightest neutralino, slepton and chargino pair production were searched for 
in the context of light gravitino models. Two possibilities
were explored: the $\tilde{\tau}_1$ NLSP  and the sleptons co-NLSP
scenarios.  No evidence for signal production was found.
Hence, the DELPHI collaboration 
sets lower limits at 95\% CL for the mass of 
the $\tilde{\chi}_1^0$ as a function of the slepton masses, 
and lower mass limits for the sleptons in all the gravitino mass range.
The limit on the chargino mass is 100\GeVcc~for all $m_{\tilde{G}}$ in the
$\tilde{\tau}_1$ NLSP scenario and 96\GeVcc~in the sleptons co-NLSP scenario.
All these results were combined to set limits on the GMSB parameter space. 
Combining all the data up to 208 GeV, a lower limit is set for the 
variable $\Lambda$\ at 17.5~TeV. Similar results have been obtained by the
ALEPH experiment concerning all these searches~\cite{aleph208}.

Mass limits for heavy stable charged particles were also derived
within the MSSM. For these particles the DELPHI collaboration sets lower mass
limits at 95\% CL for the left (right) handed stau and smuon at 97.7
(97.4)\GeVcc.  

Finally, cross-section and mass limits were derived for sgoldstinos at 95\% CL
 since no evidence of an anomalous production of events with monochromatic 
photons was observed in either of the two channels.

\subsection*{Acknowledgements}
\vskip 3 mm
 We are greatly indebted to our technical 
collaborators, to the members of the CERN-SL Division for the excellent 
performance of the LEP collider, and to the funding agencies for their
support in building and operating the DELPHI detector.\\
We acknowledge in particular the support of \\
Austrian Federal Ministry of Education, Science and Culture,
GZ 616.364/2-III/2a/98, \\
FNRS--FWO, Flanders Institute to encourage scientific and technological 
research in the industry (IWT), Belgium,  \\
FINEP, CNPq, CAPES, FUJB and FAPERJ, Brazil, \\
Czech Ministry of Industry and Trade, GA CR 202/99/1362,\\
Commission of the European Communities (DG XII), \\
Direction des Sciences de la Mati$\grave{\mbox{\rm e}}$re, CEA, France, \\
Bundesministerium f$\ddot{\mbox{\rm u}}$r Bildung, Wissenschaft, Forschung 
und Technologie, Germany,\\
General Secretariat for Research and Technology, Greece, \\
National Science Foundation (NWO) and Foundation for Research on Matter (FOM),
The Netherlands, \\
Norwegian Research Council,  \\
State Committee for Scientific Research, Poland, SPUB-M/CERN/PO3/DZ296/2000,
SPUB-M/CERN/PO3/DZ297/2000, 2P03B 104 19 and 2P03B 69 23(2002-2004)\\
JNICT--Junta Nacional de Investiga\c{c}\~{a}o Cient\'{\i}fica 
e Tecnol$\acute{\mbox{\rm o}}$gica, Portugal, \\
Vedecka grantova agentura MS SR, Slovakia, Nr. 95/5195/134, \\
Ministry of Science and Technology of the Republic of Slovenia, \\
CICYT, Spain, AEN99-0950 and AEN99-0761,  \\
The Swedish Natural Science Research Council,      \\
Particle Physics and Astronomy Research Council, UK, \\
Department of Energy, USA, DE-FG02-01ER41155. 
\newpage
%

\newpage
\clearpage

\begin{figure}[H]
\begin{center}
\epsfxsize=9.0cm \epsfysize=9.0cm \epsfbox{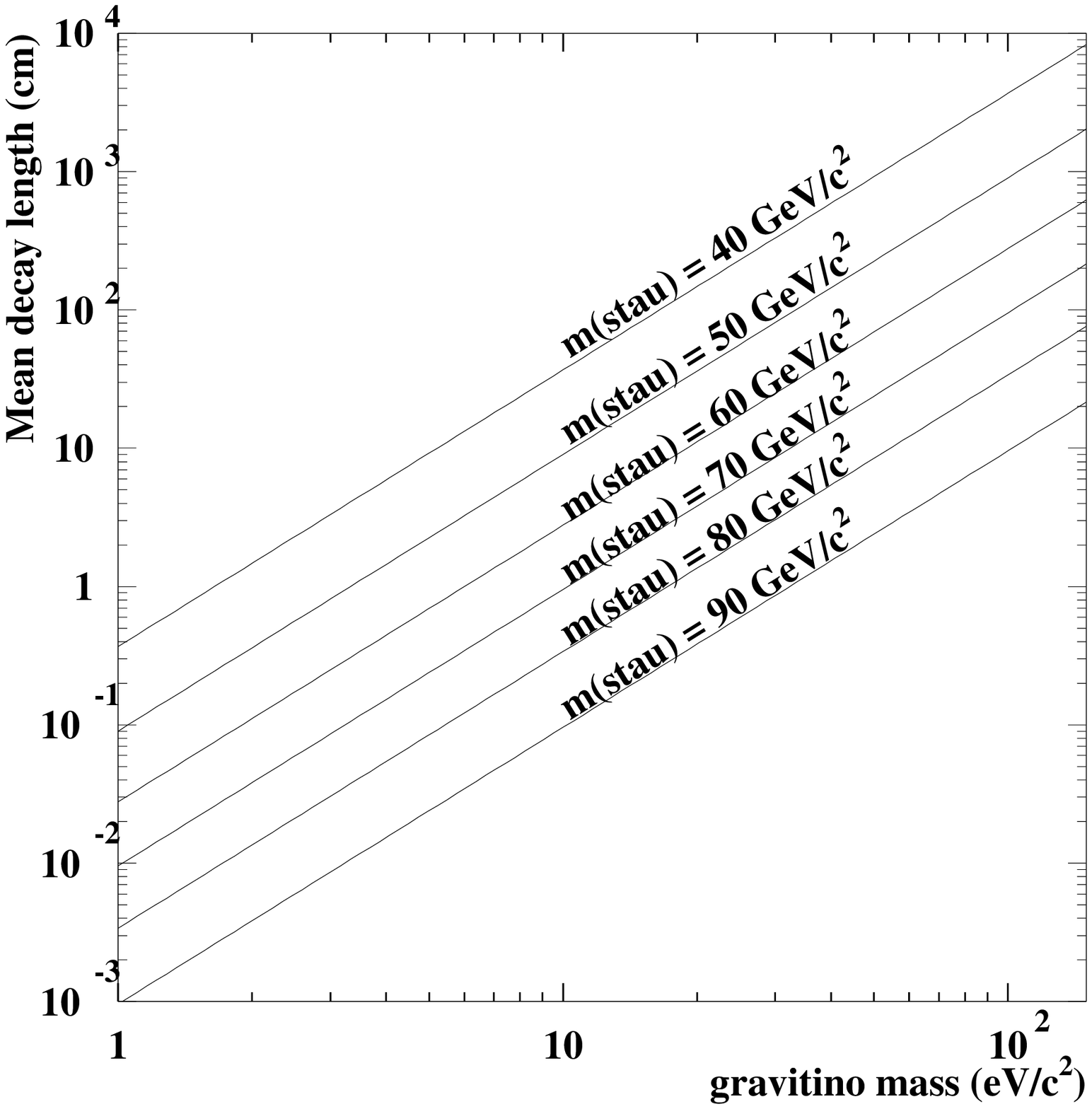}
\end{center}
\caption{\sTau~mean decay length ($\hat{L}~=~c\tau \gamma \beta$) as a function of the gravitino mass for different \sTau~mass values.}
\label{fig.meandecay}
\end{figure}
\begin{figure}[h]
\vspace{-1.5cm}
\begin{center}
\epsfxsize=7.0cm \epsfysize=7.0cm \epsfbox{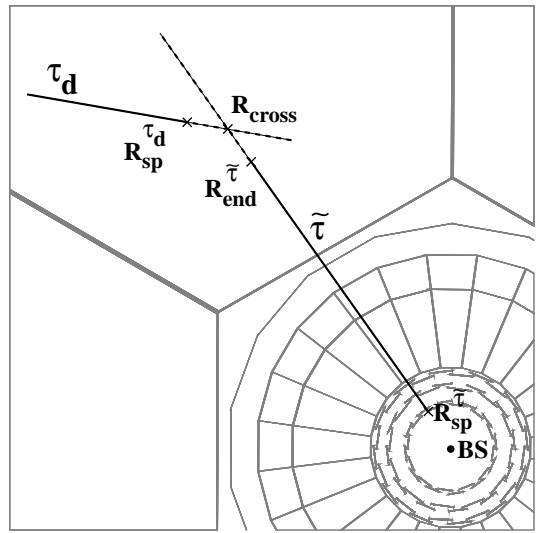}
\end{center}
\caption{Sketch illustrating the reconstruction of a secondary vertex in the plane perpendicular to the beam direction. All the radii are measured with respect to the beam spot (BS).}
\label{fig.fig-grav-def}
\end{figure}
\begin{figure}[hbpt]
\vspace{-1cm}
\centerline{\epsfxsize=16.0cm \epsfysize=13.cm \epsfbox{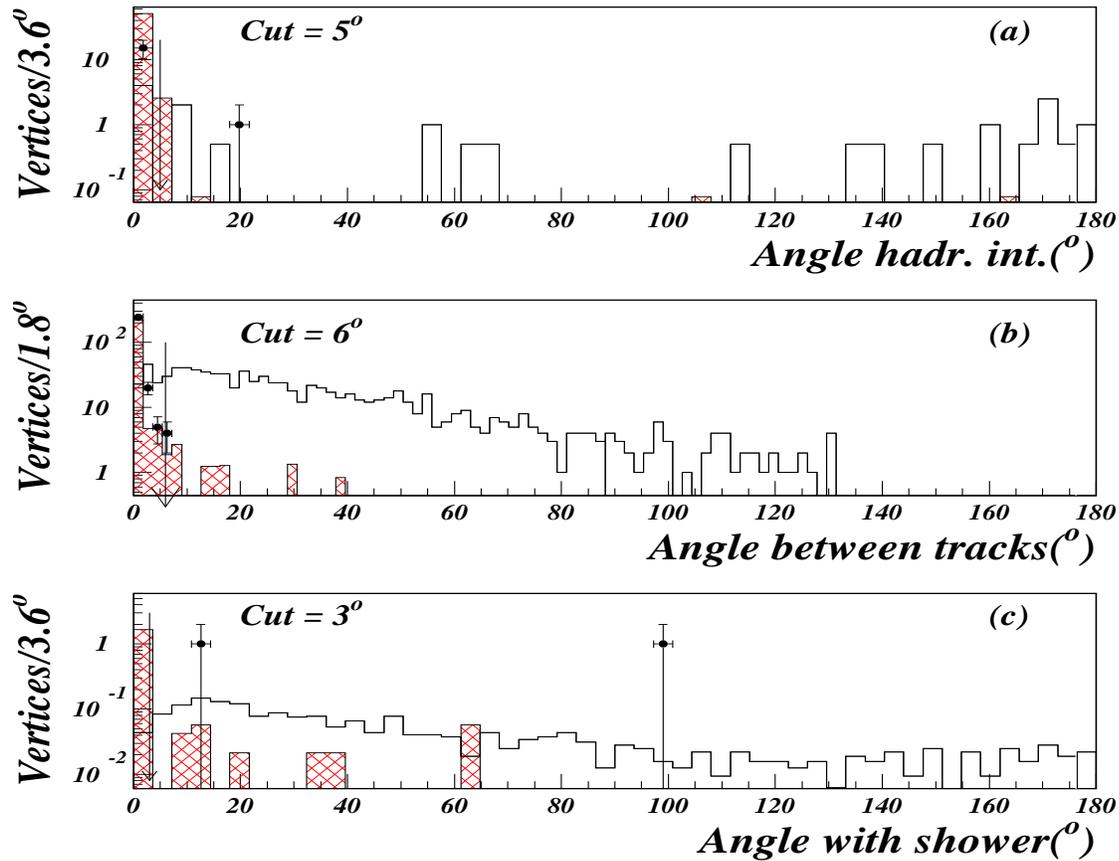}} 
\vspace{-0.5cm}
\caption[] {(a) Angle between the directions defined by the hadronic vertex
and the reconstructed vertex w.r.t. beam spot, (b) angle between the tracks
of the kink, and (c) 
angle between the electromagnetic shower and the missing momentum.
Dots are real data, cross-hatched histogram is the SM background and blank
histogram is the simulated signal ($m_{\tilde{\tau}_1} = 60$\GeVcc,
\lmean~=~50 cm and $\sqrt{s}$ = 208 GeV).}
  \label{fig:grav:kinks_BG}
\end{figure}
\begin{figure}[h]
\begin{center}
\centerline{\epsfxsize=13.0cm \epsfysize=11.0cm \epsfbox{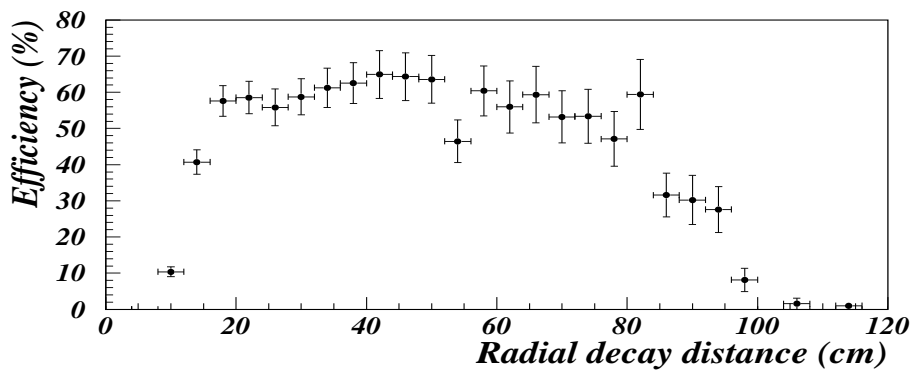}} 
\vspace{-5.cm}
\caption{Efficiency as a function of the decay radius for a sample of staus
  with $\hat{L}$ = 50 cm and $\sqrt{s}$ = 208 GeV.}
\label{fig:effivsradio}
\end{center}
\end{figure}
\begin{figure}[h]
\begin{center}
\centerline{\epsfysize=16.0cm \epsfbox{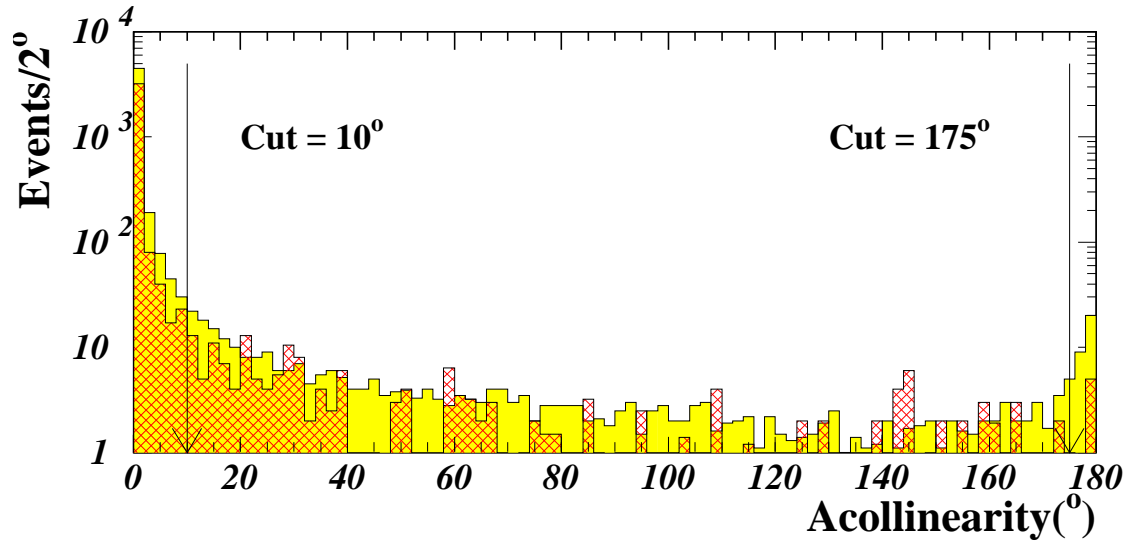}} 
\vspace{-7.0cm}
\caption{Acollinearity distribution for real data minus expected SM
  background (cross-hatched histogram), compared to cosmic ray
  events shown in dark grey. The cuts on this variable are shown with
  arrows.}
\label{fig:acoll}
\end{center}
\end{figure}
\begin{figure}[hbpt]
\vspace{-1cm}
\centerline{\epsfxsize=14.0cm \epsfysize=7.0cm \epsfbox{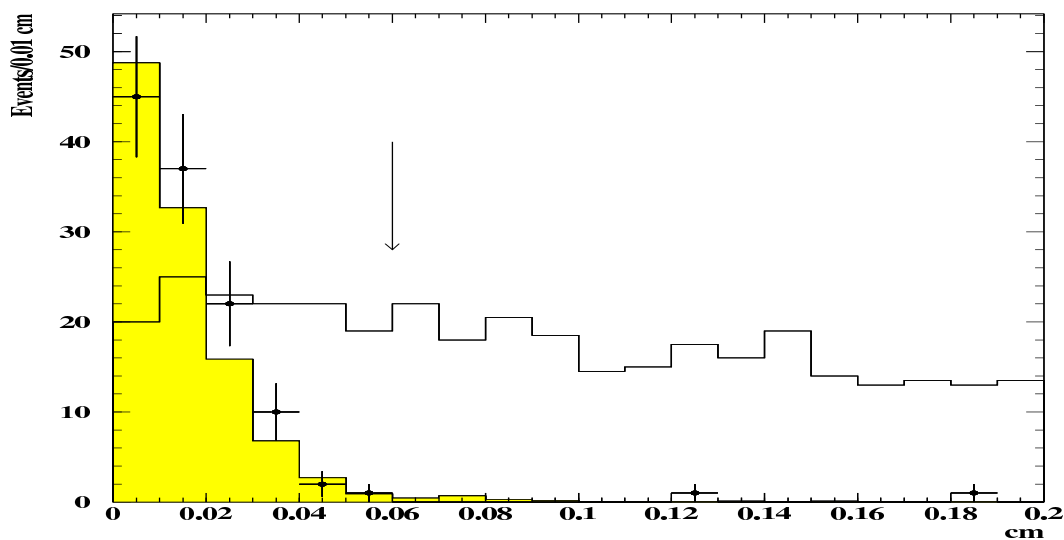}} 
\caption{$\sqrt{b_1^2 + b_2^2}$ distribution for data (dots), simulated  
SM backgrounds (grey histogram) and 
simulated signal of $m_{\tilde{\tau}_1}$ = 90 GeV/c$^2$ and
$m_{\tilde{G}}$ = 25\eVcc~at $\sqrt{s}$ = 206
GeV and
a boosted mean decay length $\sim$ 1 cm (white histogram, in arbitrary scale)
 after all other cuts applied by the small impact
parameter search.}
  \label{fig:ip-data-mc}
\end{figure}
%
\begin{figure}[htb]
\begin{center}
\mbox{\epsfxsize17.0cm 
\epsffile{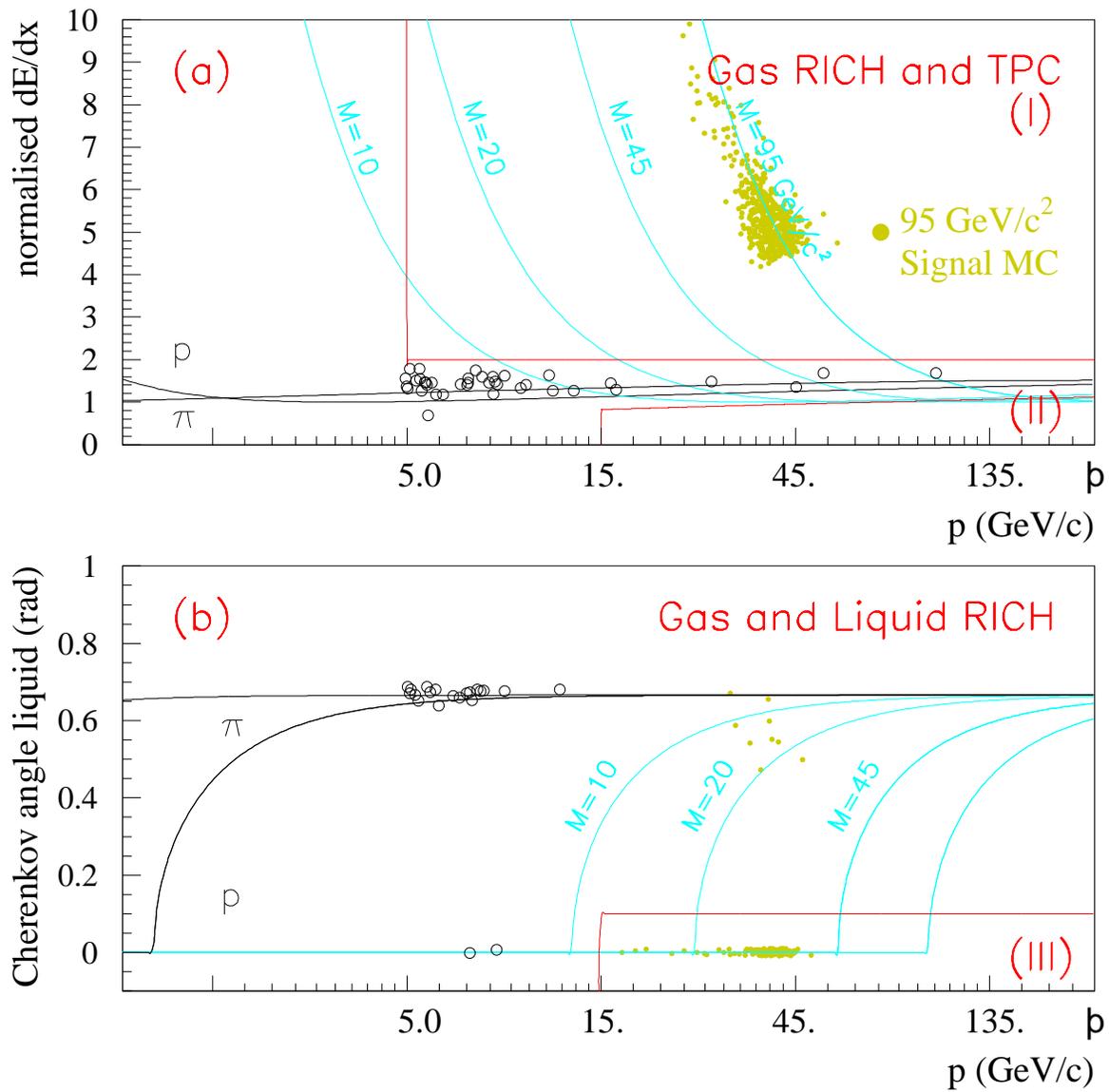} }
\end{center}
\vspace*{0.cm}
\caption{
(a) Normalised energy loss as a function of the momentum 
 after the gas veto for the 208 GeV data.
(b) Measured Cherenkov angle in the liquid radiator 
as a function of the momentum after the gas veto:
if four photons or less were observed 
in the liquid radiator, the Cherenkov angle was set equal to zero.
The areas labelled (I), (II) and (III) indicate the selection criteria
 explained in the text. Open circles are data. The small filled circles
indicate the expectation for a 95\GeVcc~mass signal with charge $\pm$e,
resulting in a large dE/dx (upper
plot) and no photons (except for a few accidental rings) in the liquid
Cherenkov counter (lower plot). The solid lines with a mass signal value
 indicate the expectation for heavy stable sleptons.}
\label{rare208}
\end{figure}
\begin{figure}[H]
\begin{center}
\begin{tabular}{c}
\epsfxsize=10.cm\epsfysize=10.cm\epsfbox{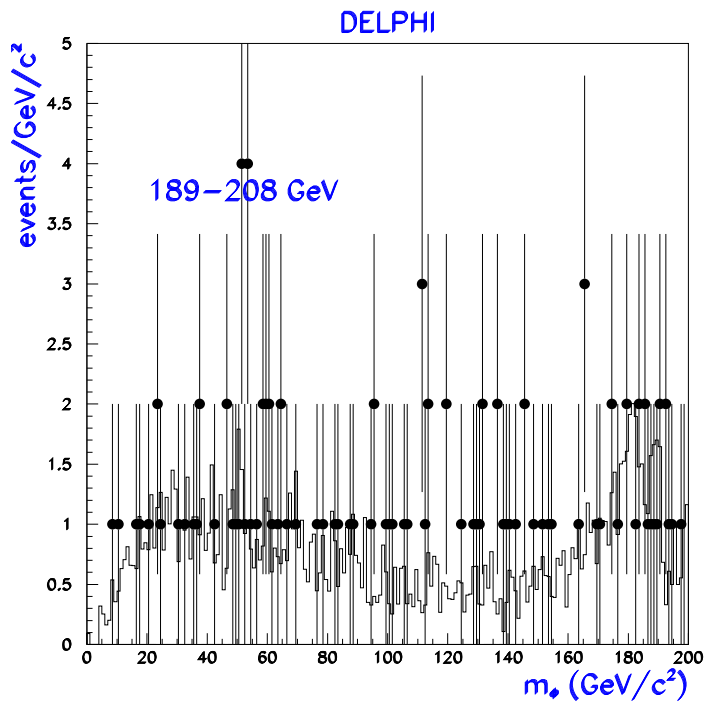} \\
a) \\
\epsfxsize=10.cm\epsfysize=10.cm\epsfbox{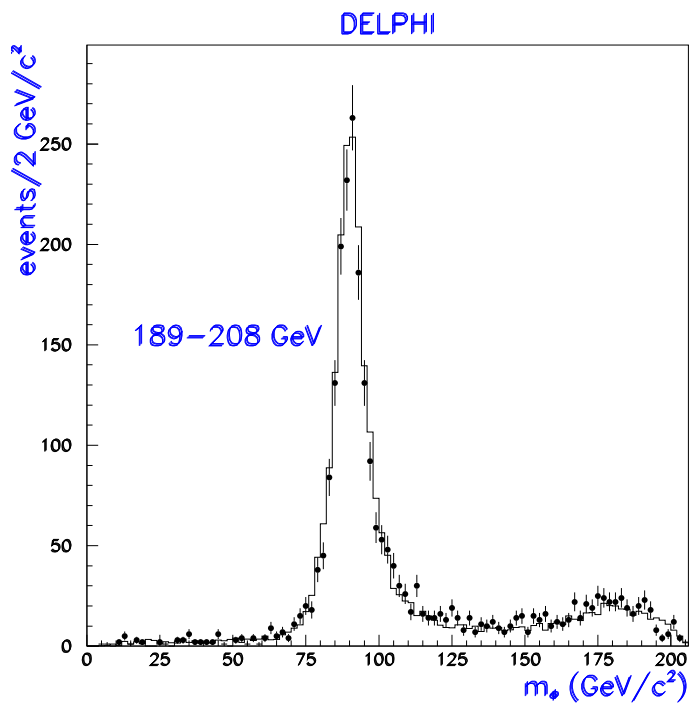} \\ 
b) \\
\end{tabular}
\end{center}
  \caption[]{a)  Photon recoil mass spectrum for 
$\gamma \gamma \gamma $ candidates (points) and the expected background
(histogram). 
The average number of entries per event in the data is 2.3. The bin size
takes into account  the experimental mass resolution 
and   the expected signal width.  b) Photon recoil mass spectrum for 
$\gamma gg $ candidates (points) and the expected background (histogram).} 
  \label{fig:msggg}
\end{figure}
\begin{figure}[H]
\begin{center}
\begin{tabular}{c}
\epsfxsize=10.cm\epsfysize=10.cm\epsfbox{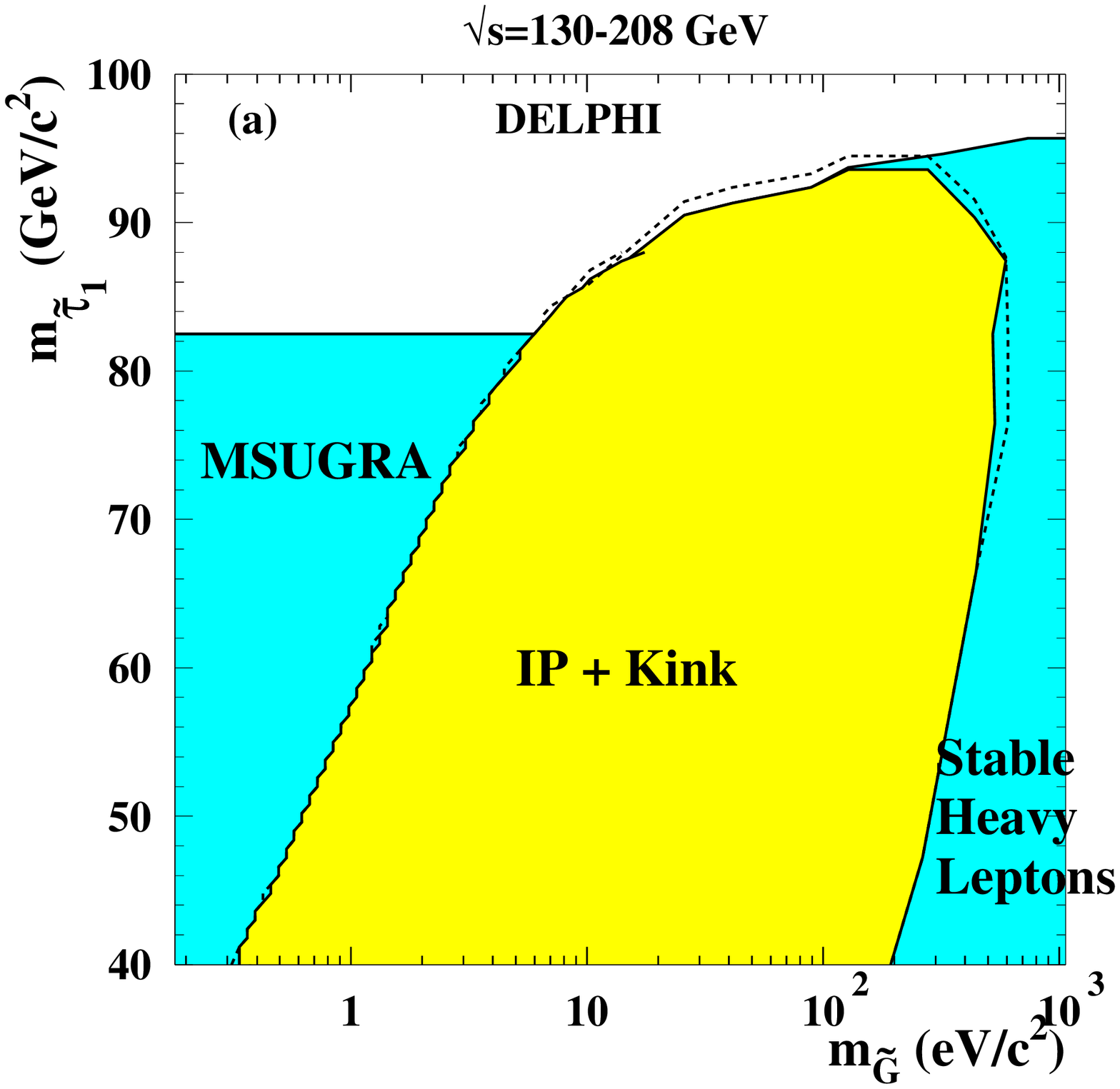} \\
\epsfxsize=10.cm\epsfysize=10.cm\epsfbox{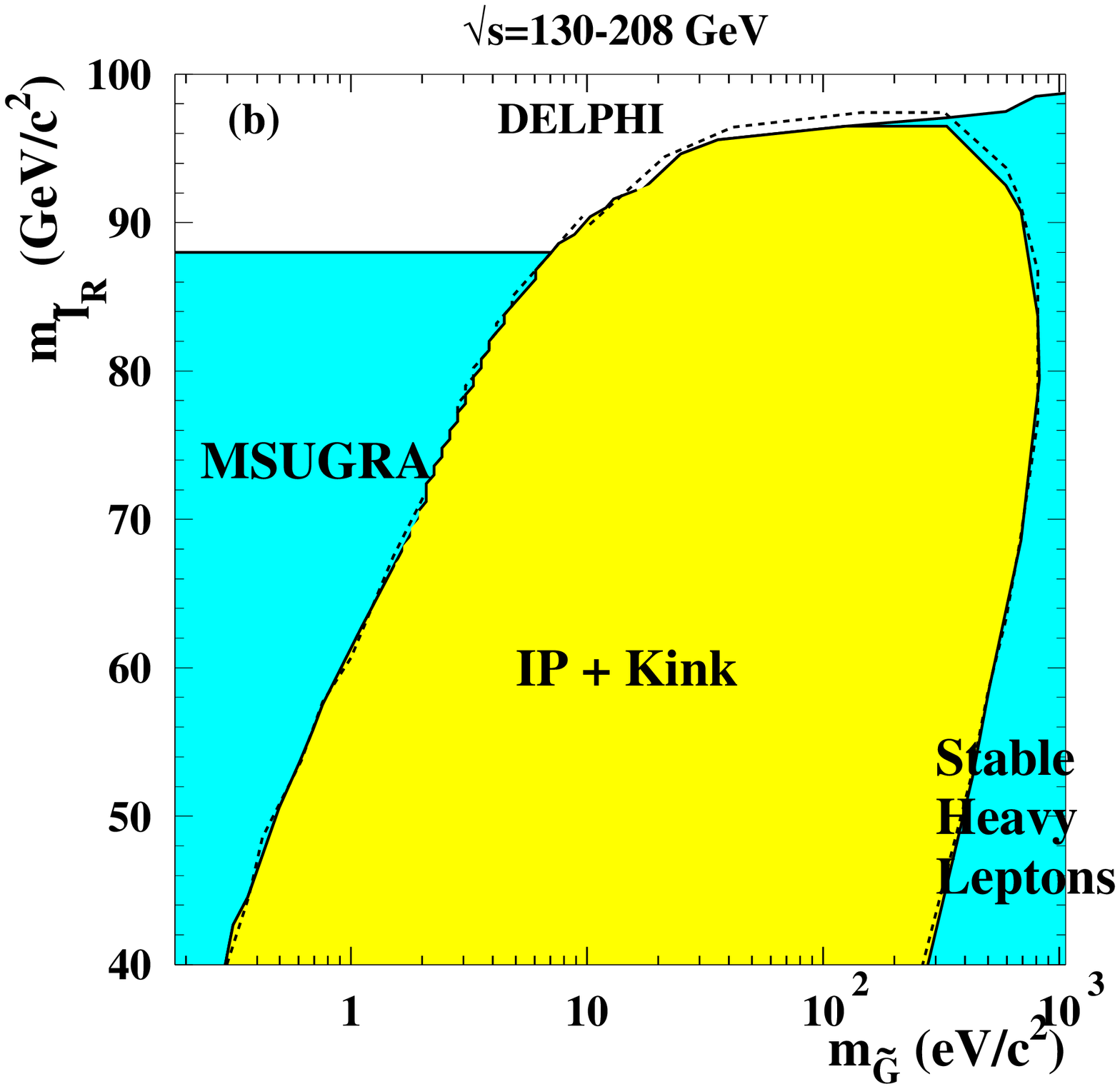} \\ 
\end{tabular}
\end{center}
  \caption[]{ 
    Exclusion regions in the 
    ($m_{\tilde{G}}$,$m_{\tilde{\tau}_1}$) (a) and
    ($m_{\tilde{G}}$,$m_{\tilde{l}_R}$) (b)
     planes at 95\%~CL for the present analyses combined 
    with the Stable Heavy
    Lepton search and the search for $\tilde{l}$ in gravity mediated
    models (MSUGRA), using all DELPHI data from 130~GeV to 208~GeV
    centre-of-mass energies. 
    The dashed line shows the expected limits for the impact parameter and
    kink searches.} 
  \label{fig:excl_208}
\end{figure}
\begin{figure}[H]
\begin{center}
\epsfxsize=15.cm\epsfysize=15.cm\epsfbox{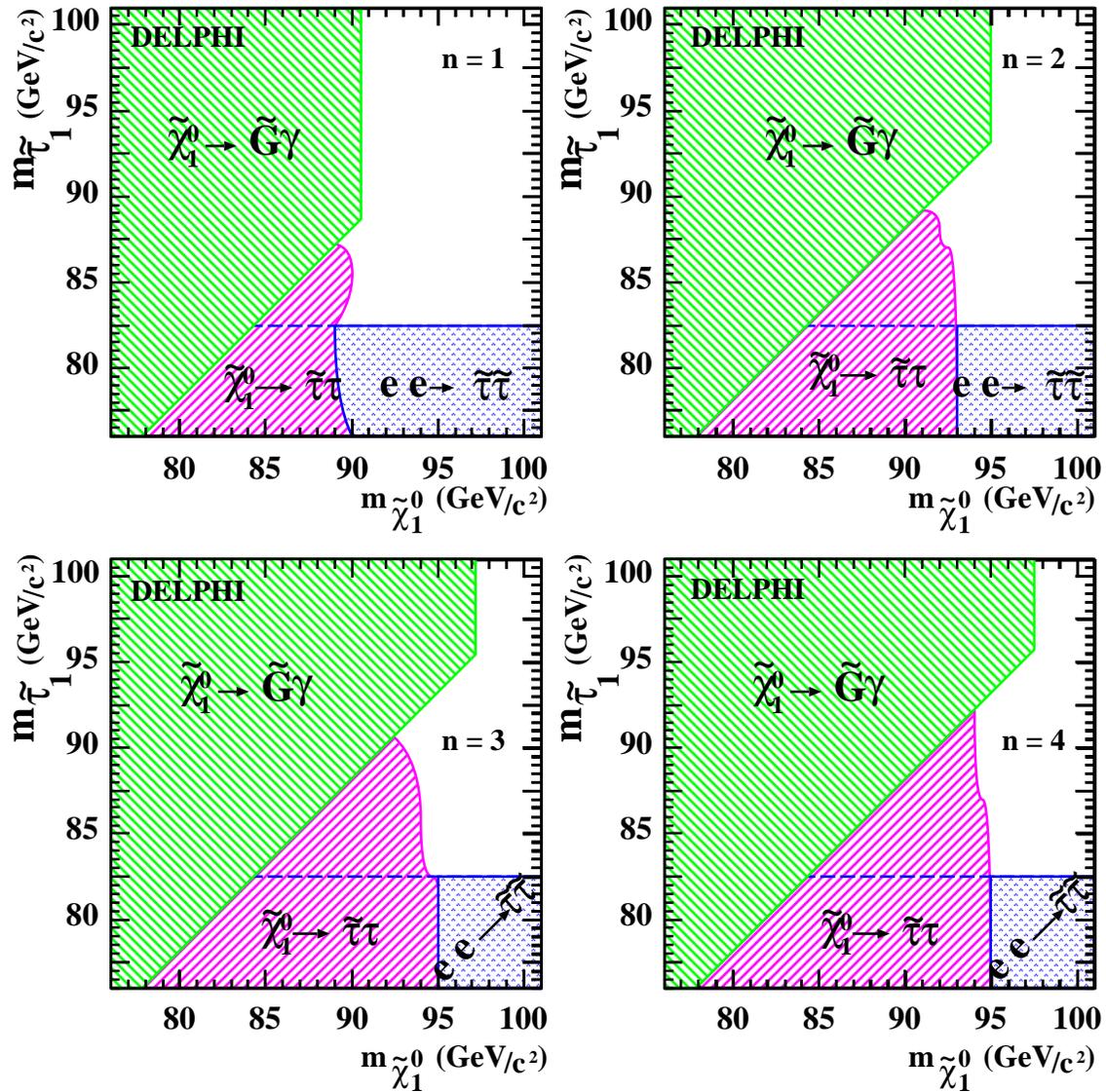} 
\end{center}
  \caption[]{Exclusion regions in the
    ($m_{\tilde{\chi}_1^0}$,$m_{\tilde{\tau}_1}$) plane at 95\%~CL for the
    neutralino analysis (positive-slope dashed area) combined with the search for neutralino pair production followed by the decay $\tilde{\chi}^0_1\rightarrow \tilde{G}\gamma$ (negative-slope dashed area), and the direct search for \staone~pair production within MSUGRA scenarios (point-hatched area), using all DELPHI data from 161~GeV to 208~GeV centre-of-mass energies.}
  \label{fig:masses}
\end{figure}
\begin{figure}[H]
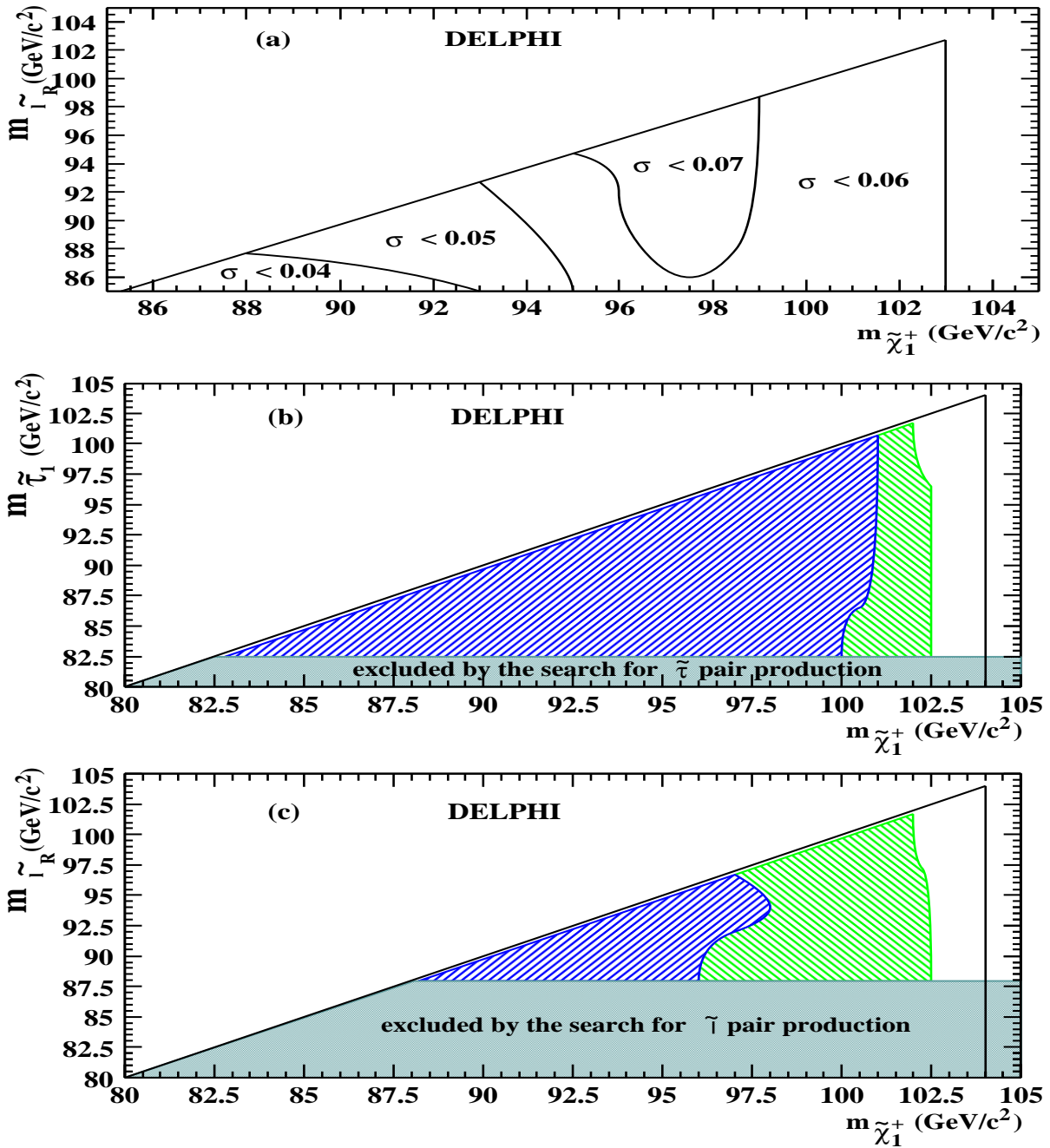

\begin{center}
\begin{tabular}{c}
\epsfxsize=16.cm\epsfysize=6.cm\epsfbox{xsec_cha_208.eps} \\
\epsfxsize=16.cm\epsfysize=12.cm\epsfbox{mass_cha_208.eps} \\ 
\end{tabular}
\end{center}
  \caption[]{(a) Limits in picobarn on the lightest chargino pair production
    cross-section at 95\% CL. Limits are shown as a function of
    $m_{\tilde{l}}$ and $m_{\tilde{\chi}_1^+}$ for $m_{\tilde{G}}$ =
    100~\eVcc. Areas excluded at 95\% CL in the
    ($m_{\tilde{\chi}_1^+}$,$m_{\tilde{\tau}_1}$) plane (b) and 
($m_{\tilde{\chi}_1^+}$,$m_{\tilde{l}_R}$) plane (c). The
    positive-slope hatched area is excluded for all gravitino masses. The  
negative-slope hatched area is only excluded \mbox{for $m_{\tilde{G}} >$
    100\eVcc}. The grey area is excluded by the search for stau pair
    production in gravity mediated supersymmetry breaking models.
Both plots have been obtained using all DELPHI data from 183~GeV to 208~GeV
    centre-of-mass energies.}  
  \label{fig:xsec-masss-chargi}
\end{figure}
\begin{figure}[htb]
\begin{center}
\epsfxsize=10.0cm\epsfysize=10.cm\centerline{\epsffile{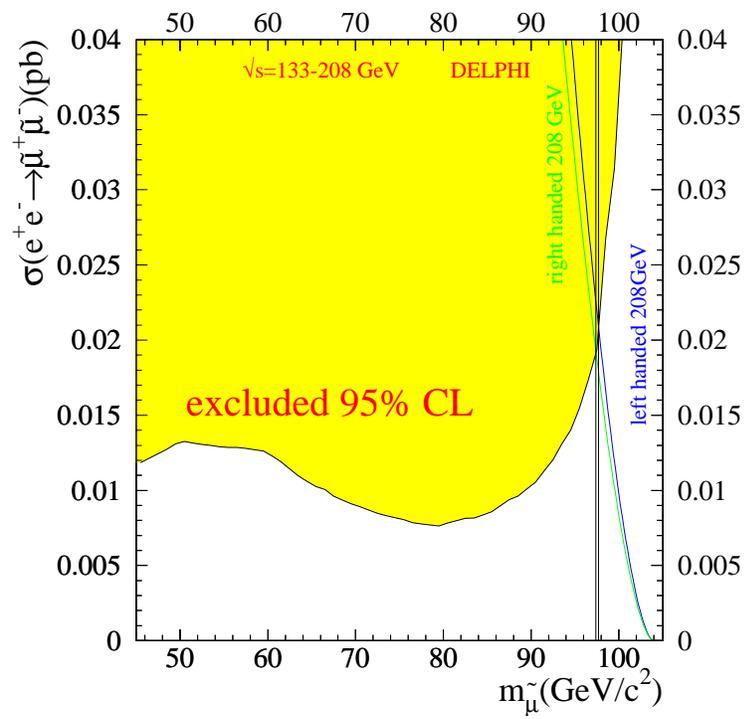}}
\end{center}
\vspace*{0.cm}
\caption{Predicted production cross-section for left and right handed
stable smuons (staus) as a function of the particle mass. The cross-section
limit indicated in the figure has been derived using all DELPHI data
between 130 and 208~GeV.}
\label{limit}
\end{figure}
\begin{figure}[th]
\begin{center}\mbox{\epsfxsize 10cm\epsfbox{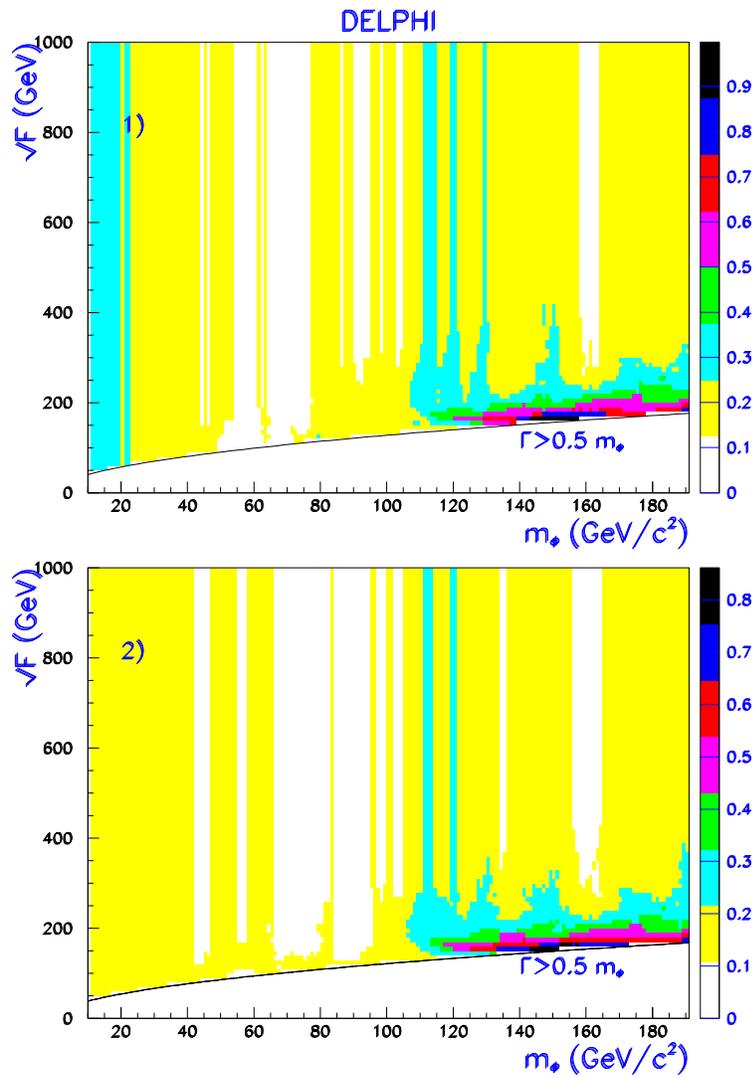}} \end{center}
\caption{
 Cross section upper limit (pb scale on the right) at the 95$\%$ CL as a
 function of $m_{\phi}$ and $\sqrt{F}$ for the two 
 sets of parameters of Table~\ref{tab:param} and using all DELPHI data
   between 189 and 208 GeV.}
\label{cslim}
\end{figure}
\begin{figure}[th]
\begin{center}\mbox{\epsfxsize 10cm\epsfbox{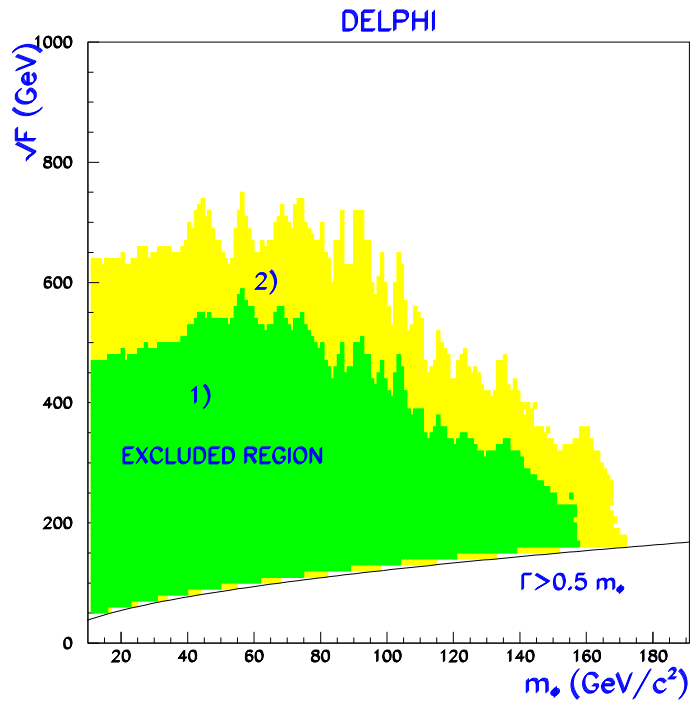}} \end{center}
\caption{
 Exclusion region at the 95$\%$ CL in the ($m_{\phi}$,$\sqrt{F}$)
 plane using all DELPHI data  between 189 and 208 GeV.
Regions 1 and 2 are 
the excluded regions once the set of parameters (labelled with 1 and 2
 respectively) of table~\ref{tab:param} are applied.}
\label{excl}
\end{figure}
\begin{figure}[htbp]
\epsfxsize=16.0cm
\centerline{\epsffile{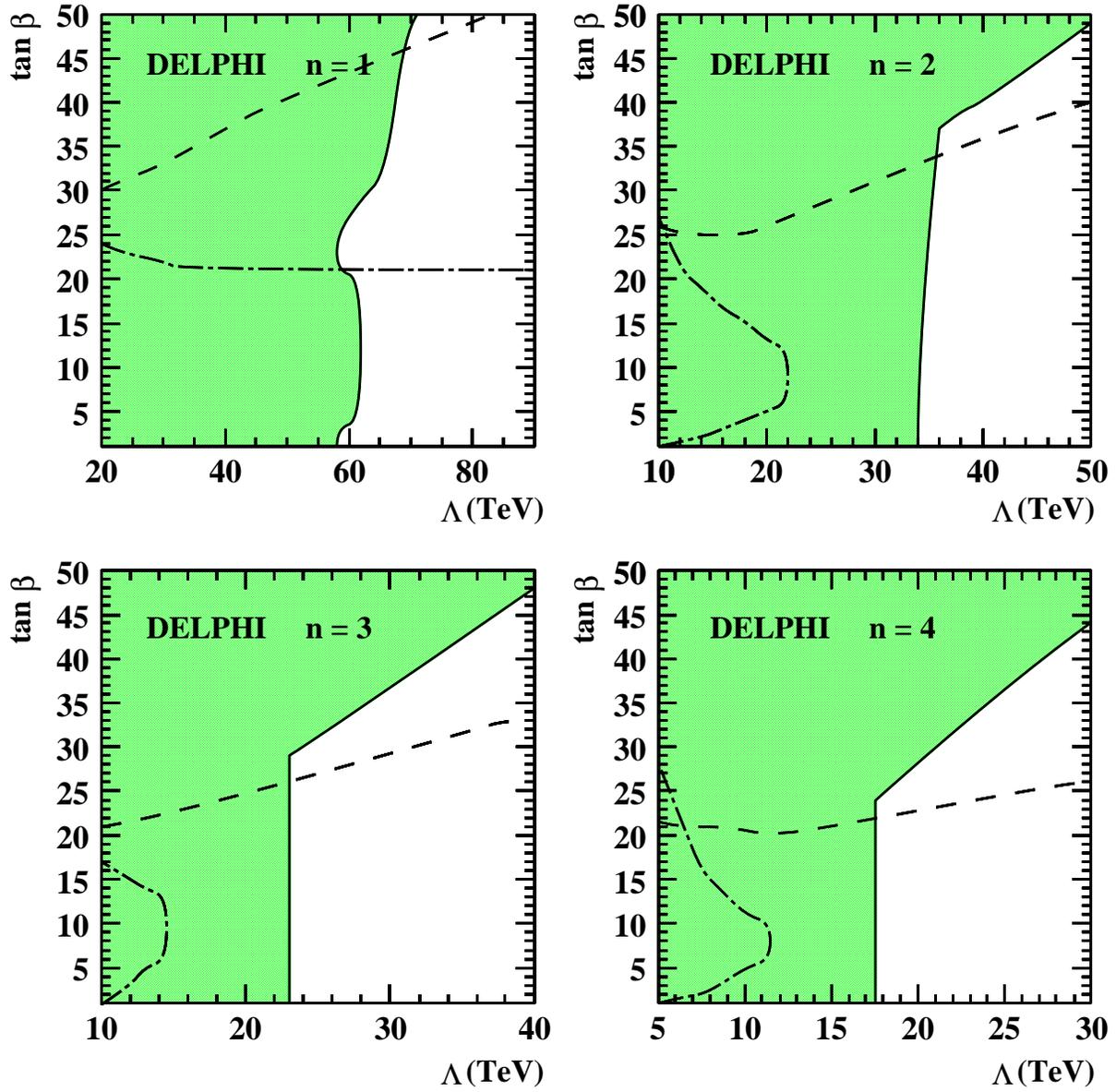}}
\caption{Shaded areas in the 
($\tan\beta , \Lambda$) plane are excluded at 95\% CL.
The areas below 
the dashed lines contain points of the GMSB parameter space with \nuno~NLSP.
The areas to the right (above for $n = 1$) of the dashed-dotted lines 
contain points of the GMSB parameter space were sleptons are the NLSP.}
\label{fig:lambda}
\end{figure}
\end{document}